\documentclass[final,3p,times,twocolumn]{elsarticle}

\usepackage{amssymb}
\usepackage{tabularx}
\usepackage{amsmath,subcaption}
\usepackage[cmintegrals]{newtxmath}
\usepackage{lineno}

\biboptions{sort&compress}
\usepackage{url,multirow,morefloats,floatflt,cancel,tfrupee}
\makeatletter
\AtBeginDocument{\@ifpackageloaded{textcomp}{}{\usepackage{textcomp}}}
\makeatother
\usepackage{colortbl}
\usepackage{xcolor}
\usepackage{pifont}
\usepackage{enumitem}  
\usepackage[nointegrals]{wasysym}
\makeatletter
\def \colone{1}
\def \coltwo{2}
\def \ncols{2}

\def \PSO{1}
\def \Yes{1}

\journal{Biomedical Signal Processing and Control}

\begin{document}

\begin{frontmatter}



\title{Predictive Modeling of Biomedical Signals Using Controlled Spatial Transformation}


\author{Jiaming~Chen, Ali~Valehi, Abolfazl~Razi\\School of Informatics, Computing, and Cyber Systems, Northern Arizona University}


\begin{abstract}
An important paradigm in smart health is developing diagnosis tools 
and monitoring a patient's heart activity through processing Electrocardiogram (ECG) signals is a key example, sue to high mortality rate of heart-related disease. However, current heart monitoring devices suffer from two important drawbacks: i) failure in capturing inter-patient variability, and ii) incapability of identifying heart abnormalities ahead of time to take effective preventive and therapeutic interventions. 

This paper proposed a novel predictive signal processing method to solve these issues. 
We propose a two-step classification framework for ECG signals, where a \textit{global classifier} recognizes severe abnormalities by comparing the signal against a universal reference model. 
The seemingly normal signals are then passed through a \textit{personalized classifier}, to recognize mild but informative signal morphology distortions. The idea is to develop a patient-specific reference for normal heart function. Another key contribution is developing a novel deviation analysis based on a controlled nonlinear transformation
\ifx \PSO \Yes
(with two computational and analytical optimization methods) 
\else
along with an analytical optimization method  
\fi
to capture significant deviations of the signal towards any of predefined 
abnormality classes. Here, we embrace the proven but overlooked fact that certain features of ECG signals reflect underlying cardiac abnormalities before the occurrences of cardiac disease. 
The proposed method achieves a classification accuracy of 96.6\% 
and provides a unique feature of \textit{predictive analysis} by providing warnings before critical heart conditions. In particular, the chance of observing a severe problem (in terms of a \textit{red} alarm) is raised by about 5\% to 10\% after observing a \textit{yellow} alarm of the same type. 
Although we used this methodology to provide early precaution messages to elderly and high-risk heart-patients, the proposed method is general and applicable to similar bio-medical signal processing applications.
\end{abstract}

\begin{keyword}
Predictive modeling\sep biomedical signal processing\sep nonlinear transformation\sep ECG signals\sep smart health


\end{keyword}

\end{frontmatter}


\section{Introduction}
\label{sec:intro}

Cardiovascular disease (CVD) is one of the leading death causes in the world, which accounts for more than 30\% of global death\citet{who}.

A common property of cardiovascular disease is the wide range of its causing factors as well as the difficulty of recognizing some important abnormalities due to the lack of painful symptoms~\citep{whooley2008depressive}. 
Indeed, some disorders are diagnosed only after they are so severe that leave the patients with little time to take therapeutic actions~\citep{circulation2010}. 
An early detection of abnormal cardiac behaviors through ECG signals prior to occurrences of CVD can save lives by enabling timely preventive actions~\cite{liew2011electrocardiogram, mandala2017ecg}. 
This is an important issue noting that sudden cardiac death (SCD), which often associates with no obvious signs and symptoms, accounts for $250,000$ to $300,000$ mortality per year in the US~\citep{SCDnumber}. These facts demand more attention by the research community to develop tools and methods for early detection of heart abnormalities. 

\subsection{ECG-based heart monitoring} \label{sec:ecg-review} 
Electrocardiogram (ECG) is the most common way of monitoring heart function. ECG signals contain abundant physiological information that reflects the heart rhythm, and the muscular and electrical status of various parts of the heart~\citep{mitdb}. 
Several computer-based automated tools have been developed for heart monitoring based on ECG signals to assist physicians and patients with more accurate diagnosis by reducing human mistakes~\citep{lagerholm2000clustering, prasad2003classification, autofs, ceylan2009novel, osowski2004support, Hu_et_al,deChazal2006,llamedo2012automatic,bbnn,ince2009generic,Kiranyaz}. 


The common spirit of these methods is processing a large dataset of annotated ECG signals and construct a reference model that facilitates evaluating test signals and classifying them into normal and different abnormality classes. These methods do not take into account the inherent variability of ECG waveform morphology among different individuals due to gender, age, body-mass index, genetic variations, and etc~\citep{intervaria}.

To address the issue of inter-patient variance, some innovative patient-specific classifiers are proposed in the last decade~\citep{Hu_et_al,deChazal2006,llamedo2012automatic,bbnn,ince2009generic,Kiranyaz}. For instance, 
Hu \textit{et al.} proposed a patient-specific classifier based on mixture of experts (MOE) by incorporating personalized annotations, provided by cardiologists, into a local classifier~\citep{Hu_et_al}. This MOE approach achieved an accuracy of 94.0\% for distinguishing \textit{ventricular} beats from other non \textit{ventricular} types.
This method enables patient-specific processing capacity but requires direct input from human experts. 
Following the design of MOE, in~\citep{deChazal2006}, an improved patient-adaptive classifier is developed by reducing the requirement of manual annotations to as few as 10 beats for training an adaptive local classifier. 

An automatic classification system is proposed in~\citep{llamedo2012automatic}, which uses experts' assistance for fine tuning, but is not fully dependent on the experts assistance. By implementing a special block-based neural network (BbNN), Jiang \textit{et al.} achieved accuracies of 98.1\% and 96.6\% in distinguishing \textit{ventricular} beats and \textit{supraventricular} beats from other types~\citep{bbnn}. The reliance of these methods on experts' annotation at any level, limits their applicability to new patients without the need for expert involvement. An important goal of the proposed methodology in this work is to enable patient adaptation without the intervention of experts. 

In~\citep{ince2009generic}, particle swarm optimization (PSO) is combined with neural network to optimize the network structure using patient-specific training data. 
Based on 1-D convolutional neural networks (CNN), a flexible algorithm is proposed in~\citep{Kiranyaz}, which adjusts its parameters using information extracted from individual records. The classifier demonstrates consistent performance over different ECG records achieving a high accuracy for distinguishing between \textit{ventricular} beats and \textit{non-Ventricular} beats with Accuracy = 98.9\%,  Sensitivity = 95.9\%, and Specificity =  99.4\%. While this approach 
does not require experts' further annotations, its performance reduces for some abnormal classes. For example, the true positive rate is 64.6\% for recognizing \textit{supraventricular} beat, and 76.1\% for recognizing \textit{fusion} beat, which is below an acceptable level.

The second and perhaps more important drawback of the existing ECG classification systems is their incapability of providing predictive alarms before the occurrence of abnormalities. Nevertheless, it has been proven by researchers that certain features of ECG signals reflect underlying cardiac abnormalities before the occurrences of cardiac disease~\cite{liew2011electrocardiogram, mandala2017ecg}. Due to the large variety of ECG parameters, the selection of optimal predictor set is a demanding task if performed manually. Our study shows that a more in-depth analysis of ECG waveforms can reveal minor abnormalities which provide hints about the upcoming severe abnormalities. Often, these minor abnormalities are considered within the normal range and hence do not used to trigger red alarms in conventional methods. To the best of our knowledge, no research work has been fully devoted for this purpose, namely the early prediction of heart abnormalities before their occurrence, which is the main focus of this project~\citep{advancewarning}.
\subsection{Predictive signal processing}   \label{sec:predictive-review} \vspace{-0.05 in}
Most existing methods treat different heart cycles independently, hence ignore the relationship between the generated labels and the upcoming abnormalities. Here, we intend to exploit these relations to improve our forecast about upcoming abnormalities. 
In this regard, we use the concept of \textit{yellow} and \textit{red} alarms and provide evidence for the fact that yellow alarms can be indicators of subsequent \textit{red} alarms. A \textit{global classifier} is trained to generate \textit{red} alarms, whereas \textit{yellow} alarms are produced through a deviation analysis, which evaluates the tendency of normal beats towards any of the predefined abnormality classes. In order to boost the performance of deviation analysis, we propose a novel controlled nonlinear transformation that maps the original feature space into a new space with a desired symmetry among abnormality clusters.

In~\citep{chen2018predictive}, a spatial transformation method is proposed based on multiple objective-particle swarm optimization (MOPSO) to realize the desired clustering symmetry. 
However, the proposed method uses an iterative optimization method with a relatively high computational complexity. Further, the interpretation of the mechanism of the system is not straightforward and easily tractable. 
The main objective of this paper is to develop a deterministic transformation with a stronger prediction power, which implements the geometry reshaping in an intuitive way. 

This paper is organized as follows. Section \ref{sec:methods} provides an overview of the entire developed methodology. The proposed deviation analysis along with the proposed controlled spatial transformation is proposed in section \ref{sec:spatial}. Section \ref{sec:result} provides numerical results to evaluate the performance of the system followed by concluding remarks in section \ref{sec:conclusion}.


\section{Methods}\label{sec:methods}
\subsection{Overview of methods} \label{sec:problem}

The main objective of this work is to solve the two issues of the current heart monitoring tools, which include i) the failure in fully capturing the inter-patient variability of ECG signals and ii) the incapability of early detection (predictive analysis). 

The essence of most existing methods is developing a reference model by pooling ECG beats from different patients together, to be used for all new patients. This approach ignores the inter-patient variation even among signal waveforms belonging to the same abnormality class. 

Our proposed method comprises two stages, where the first stage implements a \textit{global classifier} which follows the conventional approach of identify severe abnormalities in terms of \textit{red} alarms. The second step offers a subsequent \textit{personalized classifier} that takes into account the patient-specific signal morphological variations as presented in section \ref{sec:spatial}. The idea is to use \textit{normal} samples from a patient to develop a \textit{personalized normal range} for each patient as a reference to identify the occurrence and severity of deviations in the subsequent signal samples. The type of abnormalities, however, are determined using abnormal samples gathered from different patients. Therefore, we are able to recognize specific abnormalities even if we have no prior reference for that specific abnormality for the new patient. The \textit{personalized normal range} is developed and refined gradually during the lifetime of the device.

\begin{figure}[t]
\centering
\ifx \ncols \coltwo
\includegraphics[width = \columnwidth]{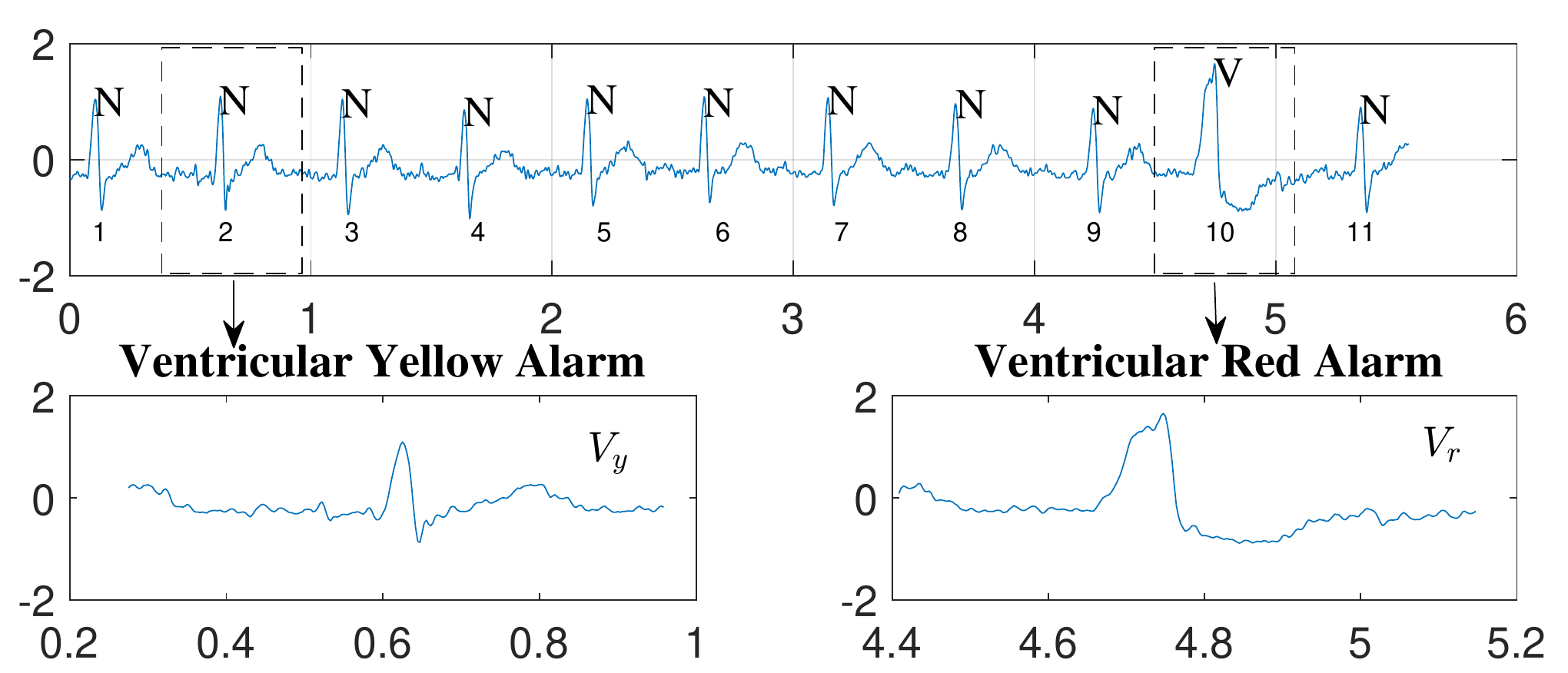}
\else
\includegraphics[width = 0.8\columnwidth]{figures/predicting_record215_0719.pdf}
\fi
\caption{A yellow alarm represents a minor abnormality (left: beat 2) which can be indicator of an upcoming severe abnormality of the same type in terms of a \textit{red} alarm (right: beats 10).}
\label{fig:y-r}
\end{figure}

This approach enables another useful feature of \textit{predictive analysis} by providing warning messages in terms of \textit{yellow} alarms before the occurrence of severe heart abnormalities. 
Fig.\ref{fig:y-r} illustrates this concept with a sample ECG recording, record \#215 in the MIT-BIH Arrhythmia database (MITDB) dataset~\citep{mitdb}. In Fig.\ref{fig:y-r} the $10^{th}$ beat, represents an abnormality of type \textit{venticular} (shown by $V_r$). The signal morphology exhibits a flattened T wave, which is severe enough to be captured by the \textit{global classifier} using a universal model. Other beats are labeled as normal by the trained global classifier. However, a more close investigation of the preceding beats reveals that the $2^{nd}$ beat includes a similar but milder distortion. These mild abnormalities are typically ignored by conventional methods to avoid calling false alarms. 
Here, we develop a \textit{personalized classifier} which enables us to catch these mild abnormalities, quantify their level of deviation, and trigger informative \textit{yellow alarms} ahead of time.

\subsection{Utilized ECG dataset} \label{sec:data}
Following the recommendation by the Association for the Advancement of Medical Instrumentation (AAMI)~\citep{aami} for reporting ECG classifier performance, we use the ECG signals in the MITDB database~\citep{mitdb} acquired from Physionet~\citep{physionet}. 
It includes 48 records collected from 47 individuals. Each record contains two channels of ECG signals along with annotations for each cardiac cycle. We use the ECG signal obtained from the \textit{MLII} channel, which is more informative and commonly used in automated ECG analysis systems~\citep{karpagachelvi2010ecg}.
The original labels include 16 types. However, according to AAMI, they are further grouped into 4 major classes: class N (normal and bundle branch block beat types), class V (\textit{Ventricular} type), class S (\textit{Supraventricular} type) and class F (\textit{Fusion} of normal and ventricular types). The class Q, which includes unclassified and paced beats is discarded due to the limited number of samples (only four records). For the purpose of training and evaluating classifiers, MITDB is split into training dataset (DS1) and test dataset (DS2), each containing 22 records with a balanced number for four classes~\citep{autofs}. 

\subsection{ECG signal processing} \label{sec:proc}

The data preparation for ECG signal in this paper comprises the following steps:\\ \noindent \textbf{Signal preprocessing:} a band-pass filter using \textit{Daubechies} wavelet is applied to remove the electromyogram noise and the baseline wander from the ECG signal following similar works, e.g.~\citep{denoise}. 

\noindent \textbf{Delineation:} Each beat in a ECG signal contains five fiducial peaks: P, Q, R, S and T. A wavelet-based delineation method, proposed in~\citep{2012qrs}, is used to detect the five peaks respectively, which enables splitting the signals into cardiac cycles (or beats).

\noindent \textbf{Segmentation:} The ECG signal may include transient terms that negatively impact the classification results. In order to smooth out these terms, we combine $s_{w}\geq 1$ consecutive cardiac cycles into a single \textit{segment}. We can arbitrarily slide a segment $n_{s}$ cycles forward to generate a new segment, where $1 \leq n_{s}\leq s_{w}$ to utilize all cycles. In this paper, we choose non-overlapping segmentation with $n_{s}=s_{w}=3$, which yields the best classification accuracy~\citep{jchen}.

\noindent \textbf{Feature extraction:} ECG signals typically are within the 5 Hz to 15 Hz frequency range. However, according to several studies, the signal power spectral density shows significant differences for various signal classes~\citep{martinez2004wavelet}. Likewise, some other temporal features (such as the duration between the Q and T waves, the ratio of P wave to R wave, etc.) present different levels of correlations with specific signal classes~\citep{autofs}. Therefore, we choose to use a combination of temporal, morphological and spectral features as detailed in Table \ref{table:features}. 
Moreover, to account for segment-level characteristics as well as cycle-level properties, the extracted features include both cycle-based features (SET 1) and segment-based features (SET 2). SET1 includes the average and standard deviation of $8$ features for the $s_w=3$ cardiac cycles within a segment, and SET2 contains $6$ features representing the overall characteristics of the signal within a segment, so it is calculated only once per segment. Therefore, we have a total of $8 \times 2 + 6 =22$ features per segment as shown in Table \ref{table:features}. 

\ifx \ncols \coltwo 
\begin{table}[t]
\footnotesize
\caption{Features extracted from ECG signals}
\label{table:features}
\centering
\begin{tabular}{|m{12em}|m{12em}|}
\hline
SET1 & SET2 \\ \hline
Temporal Features: QRS duration, QT duration, PR duration & Temporal Features: mean $(R_{i+1}-R_i)$, mean $(R_i-R_{avg})$ \\ \hline
Morphological Features: max positive peak to second peak ratio & Morphological Features: signal average energy, max positive peak, max negative peak, peak to energy ratio \\ \hline
Frequency Domain Features: signal power level at 7.5Hz, 10Hz, 12.5Hz, 15Hz &  \\ \hline
\end{tabular}
\end{table}
\else
\begin{table}[t]
\footnotesize
\caption{Features extracted from ECG signals}
\label{table:features}
\centering
\begin{tabular}{|m{15em}|m{15em}|}
\hline
SET1 & SET2 \\ \hline
Temporal Features: QRS duration, QT duration, PR duration & Temporal Features: mean$(R_{i+1}-R_i)$, mean$(R_i-R_{avg})$ \\ \hline
Morphological Features: max positive peak to second peak ratio &  Morphological Features: signal average energy, max positive peak, max negative peak, peak to energy ratio \\ \hline
 Frequency Domain Features: signal power level at 7.5Hz, 10Hz, 12.5Hz, 15Hz &  \\ \hline
\end{tabular}
\end{table} 	
\fi

\noindent \textbf{Dimensionality reduction:} In order to achieve a better classification performance and eliminate potentially overlapping features, we use \textit{principal component analysis} (PCA) to map the feature vectors $\mathbf{x}_k$ from the original 22-dimensional space $\Omega^{22}$ into an 8-dimensional feature space, $\Omega^8$. Dimensionality reduction methods such as PCA are preferred over explicit feature selection methods in biomedical signal processing, since they retain more information~\citep{martis2013ecg}. 

For each segment, a new annotation is generated by integrating the annotations of member beats within the segment. If all beats are labeled as normal, the segment is also labeled \textit{normal}. Otherwise, if the segment includes some beats with abnormality types of the same type, the segment is labeled as \textit{abnormal} of that type. However, if beats of more than one abnormality types present in a segment, it is 
discarded. For instance, if the $w_s=3$ member beats of the $k^{th}$ segment are labeled as "$NNN$", "$VNV$", "$SSS$", this segment is respectively labeled as $y_k=N$, $y_k=V$, and $y_k=S$, where $y_k$ 
is the true label for sample $\mathbf{x}_k$; whereas a segment with member beats labeled as "$NSV$" is discarded. 
The number of resulting data samples after segmentation and annotation is presented in Table \ref{table:ds}. 

\begin{table}[h]
\footnotesize
	\centering
	\caption{Training and test datasets obtained using MITDB dataset.}
	\begin{tabular}{|m{5em}|c|c|c|c|c|}
		\hline 
		& \multicolumn{4}{c}{Number of segments per AAMI class} &\\ 
		\hline 
		Evaluation Dataset& N & V & S & F &Total \\ 
		\hline 
		DS1:Training & 11721& 2356 & 862 & 256 & 15195\\ 
		\hline 
		DS2:Test & 12633 & 2053 & 550 & 121 & 15357 \\ 
		\hline 
		Total & 24354 & 4409 & 1412 & 377 & 30552 \\ 
		\hline 
	\end{tabular}
	\label{table:ds} 
\end{table}

\subsection{Classification Framework}\label{sec:classification}

Based on our previous work~\cite{chen2018predictive} and other similar works~\cite{Hu_et_al,deChazal2006,llamedo2012automatic}, here we propose a two-staged classification structure which includes \textit{global} and \textit{personalized} classifiers, as depicted in Fig.\ref{fig:flow}. 

The \textit{global classifier} is trained using the whole training set to process test signals and identify beats with severe morbidity. Depending on the application, different classification algorithms can be utilized~\cite{llamedo2012automatic}. Here, we use \textit{k-nearest neighbors} (kNN) algorithm with $k=10$, for its superior performance with an acceptable complexity based on our simulations. The feature space $\Omega^d$ is partitioned into $C$ clusters, i.e. $\Omega^{d}=\{\mathcal{X}_0,\mathcal{X}_1,\dots,\mathcal{X}_{C-1}\}$, where $\mathcal{X}_0$ represents the normal cluster and $\mathcal{X}_1,\dots,\mathcal{X}_{C-1}$ are the abnormality clusters. In this work, we have $C=4$, $d=8$,  $\Omega^{8}=\{\mathcal{N},\mathcal{S},\mathcal{V},\mathcal{F}\}$ and the predicted labels are shown as $\hat{y}_k \in \mathcal{Y}=\{N_r,V_r,S_r,F_r\}$.

An abnormal label of any type $\hat{y}_k \in \{V_r,S_r,F_r\}$ generated by the \textit{global classifier} is considered a \textit{red alarm}. However, the samples classified as normal by the \textit{global classifier} ($N_r$) undergo a subsequent analysis is required to process normal samples ($N_r$). The main goal of \textit{personalized classifier} is to develop a new set of decision rules to determine whether or not the mild deviations in the normal samples are worthy of calling yellow alarms. 


\ifx \ncols \coltwo 
\begin{figure}[thpb]
\centering
\includegraphics[scale=.42]{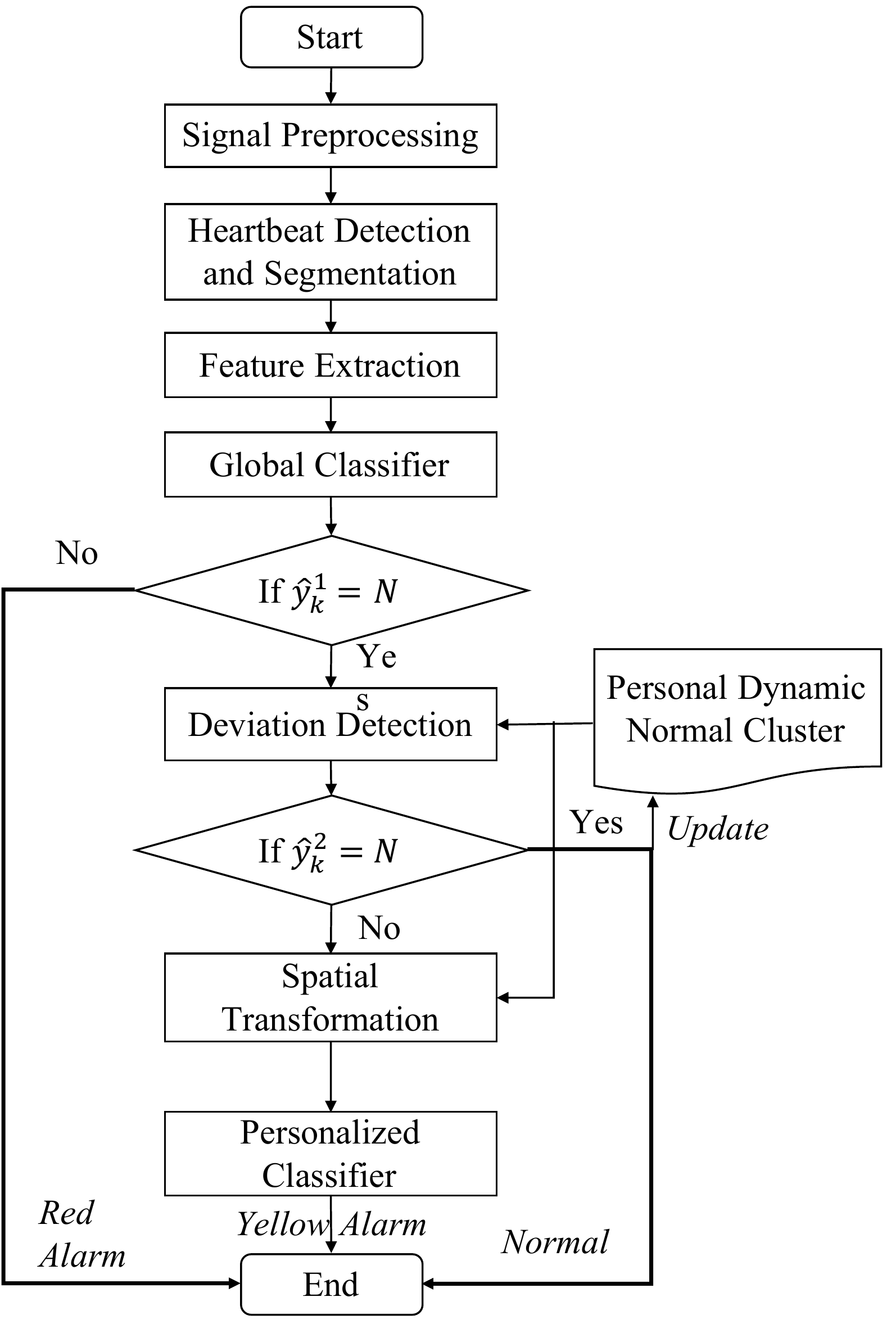}
\caption{The general flowchart of proposed framework}
\label{fig:flow}
\end{figure}
\else
\begin{figure}[thpb]
\centering
\includegraphics[scale=.4]{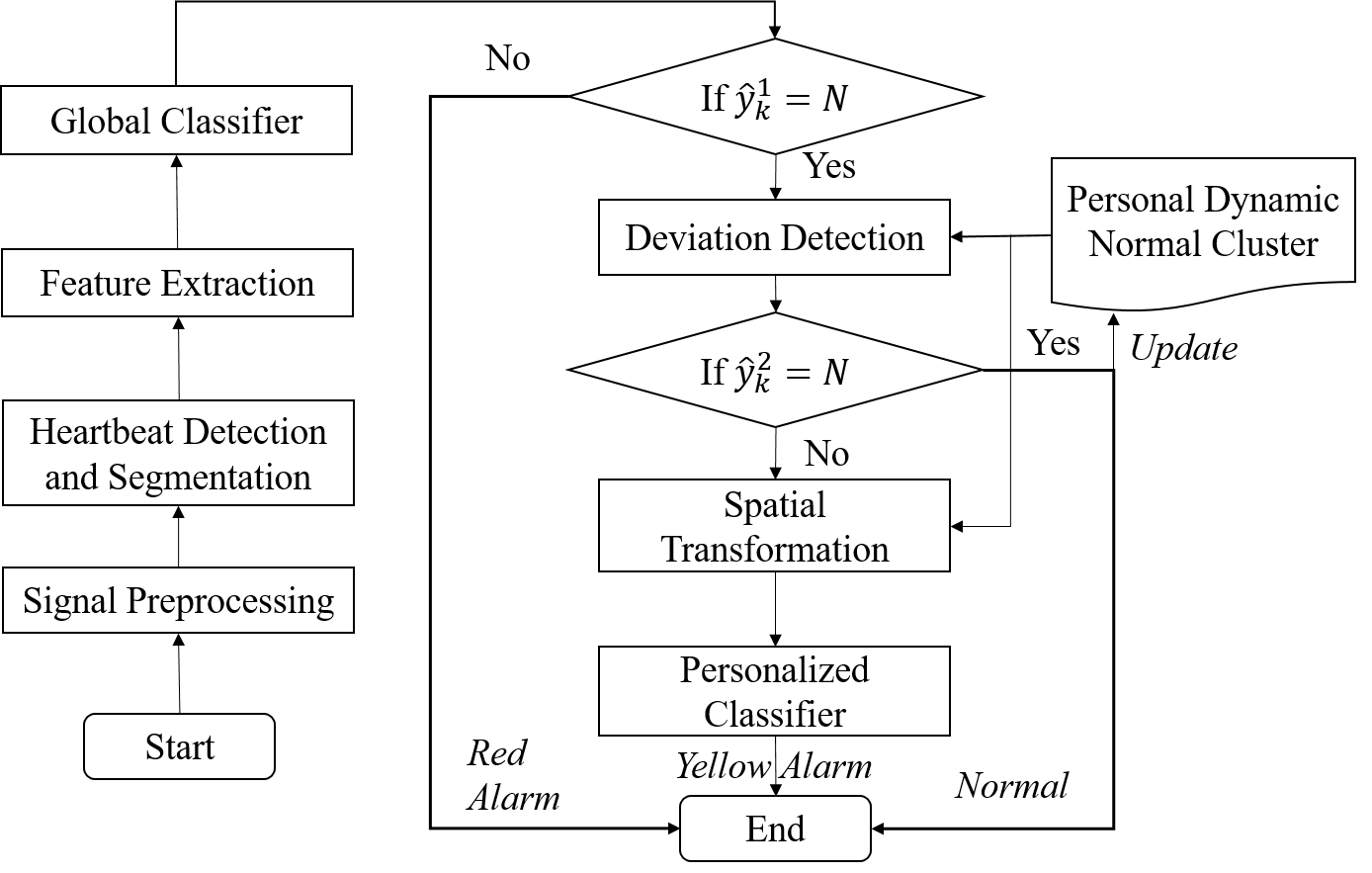}
\caption{The general flowchart of proposed framework}
\label{fig:flow}
\end{figure}
\fi

In order to develop a reference for patient-specific normal functionality for the \textit{personalized classifier}, normal beats within the first 5 minutes of the ECG signal for patient $i$ are selected as the initialization set for his/her \textit{personalized dynamic normal cluster} at time $t$, denoted by $\mathcal{N}_t(i)$. As we collect more samples from the patient, the normally labeled samples within the last 5 minutes are used to update $\mathcal{N}_t(i)=\{\mathbf{x}_k(i): y_k(i)=N_r \text{ for } k=t,t-1,t-2,\cdots\}.$ 

To distinguish between the firm normal and fuzzy states, we use a binary classifier for normal $\{N\}$ versus abnormal $\{S,V,F\}$. We firstly calculate the following distance metrics:
\begin{align}
\label{eq:matrices}
\nonumber
&R_i^{\max}=\underset{\mathbf{x}_j\in\mathcal{N}_k(i),\mathbf{x}_k\in\mathcal{N}_k(i)}{\max}\{\sqrt{(\mathbf{x}_j-\mathbf{x}_k)^2}\},
\\
\nonumber
&D_\mathcal{X}(\mathbf{x}_k(i))=\underset{\mathbf{x} \in\mathcal{X}}{\text{median}}\{\sqrt{(\mathbf{x}_k(i)-\mathbf{x})^2}\},
\\
&D_\mathcal{N}^{\max}(\mathbf{x}_k(i))=\underset{\mathbf{x} \in\mathcal{N}_k(i)}{\text{max}}\{\sqrt{(\mathbf{x}_k(i)-\mathbf{x})^2}\},
\end{align}

where $R_i^{\max}$ is the approximate diameter of the updated \textit{personalized normal cluster} for user i. Likewise, $D_\mathcal{X}(\mathbf{x}_k(i))$  is the median of distances from the current sample $\mathbf{x}_k(i)$ to samples in cluster $\mathcal{X}$, where $\mathcal{X}$ is any of the the abnormality clusters: ${\mathcal{S},\mathcal{V},\mathcal{F}}$. Finally, $D_\mathcal{N}^{\max}(\mathbf{x}_k(i))$, is the maximum distance of the current sample from the recently aggregated normal samples for patient $i$. 
Then, the following conditions in (\ref{eq:condition1}) are examined for new samples: 
\begin{align}\label{eq:condition1}
&D_\mathcal{N}^{\max}(\mathbf{x}_k(i))  \leq\alpha R_i^{\max},\\
\label{eq:condition2}
&D_\mathcal{N}(\mathbf{x}_k(i)) < D_\mathcal{X}(\mathbf{x}_k(i)) &\text{for~} \mathcal{X}\in\{\mathcal{S},\mathcal{V},\mathcal{F}\}, 
\end{align}
where (\ref{eq:condition1}) verifies if the deviation of the sample is within a range defined by $\alpha$, and (\ref{eq:condition2}) verifies that the maximum distance of the current sample from the normal samples is lower than its maximum distance from abnormal samples of any type.

If a normally labeled sample by the \textit{global classifier} ($\hat{y}_k=N_r$)
, is again confirmed as normal, it is labeled as $\hat{y}_k=N$ 
and is used to update $\mathcal{N}_k(i)$. Otherwise, the system 
passes it to the subsequent \textit{personalized classifier}. The \textit{personalized classifier} uses controlled transformation with optimized parameters to discern the deviation to different morbid types regardless of the clustering topology within the original feature space, as detailed in section \ref{sec:spatial}.
After performing both \textit{global} and \textit{personalized} classification steps, each sample $\mathbf{x}_k(i)$ is mapped to a label $\hat{y}_k \in \{N,V_y,S_y,F_y,V_r,S_r,F_r\}$, where $N$ stands for the normal state, and $X_y$ and $X_r$ stand for \textit{yellow} and \textit{red} alarms of type $X$.

Note that the abnormality clusters ${\mathcal{S},\mathcal{V},\mathcal{F}}$ are developed based on the entire training dataset DS1. 
On the other hand, the \textit{personalized} classifier constantly refines $\mathcal{N}_k(i)$ over time based on the most recently accumulated normal samples
, so no expert annotation is required for new samples. 

\section{Spatial Transformation} 
\label{sec:spatial}

\begin{figure}[h]
\centering
\ifx \ncols \coltwo
\includegraphics[width=0.99\columnwidth]{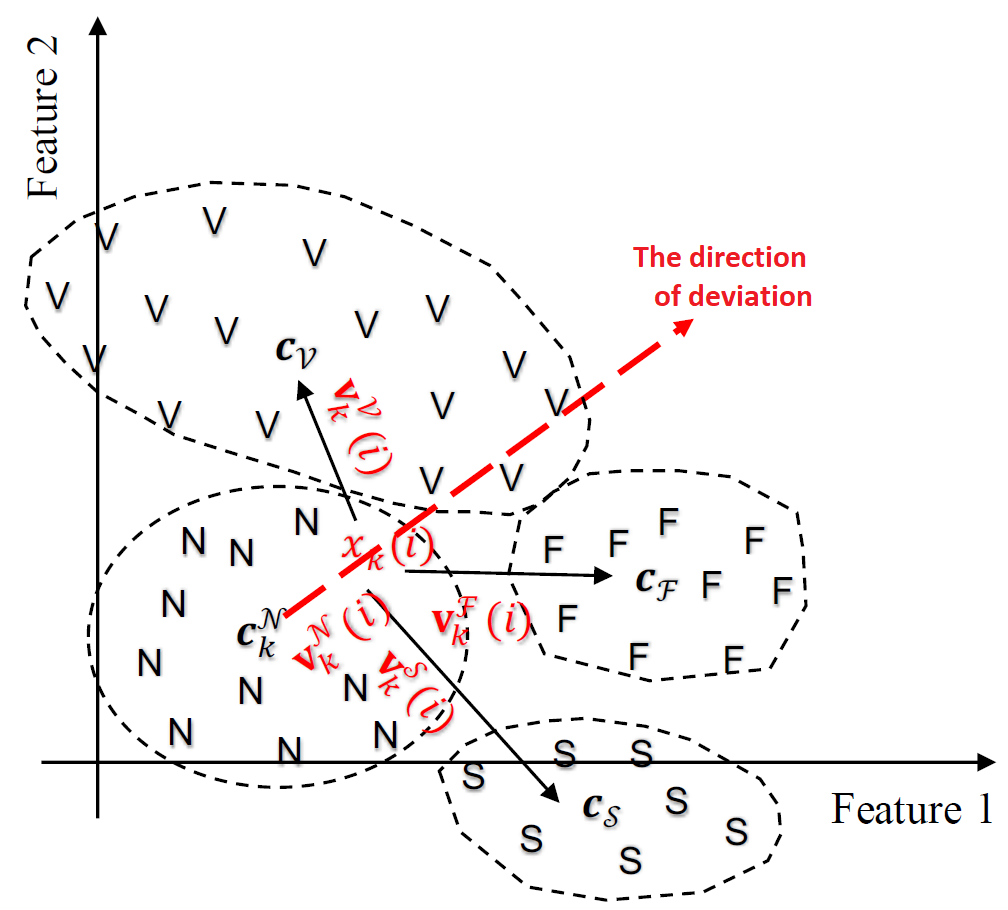}
\else
\includegraphics[width=0.69\columnwidth]{figures/cluster_2D_2.png}
\fi
\caption{Representation of vectors $\mathbf{v}_k^\mathcal{N}(i)$, $\mathbf{v}_k^\mathcal{F}(i)$, $\mathbf{v}_k^\mathcal{V}(i)$, and $\mathbf{v}_k^\mathcal{F}(i)$ used to quantify the deviation of current normal sample to one of the abnormality classes.} 
\label{fig:vecs}
\end{figure}

Deviation analysis is performed by observing the relative distance of current sample in the feature space with respect to the patient's normal cluster as well as the global abnormality clusters. Therefore, it is vital to obtain a symmetric geometry among abnormality clusters to avoid bias to the topological characteristics of the training dataset. 
A natural choice for deviation analysis in the high-dimensional space, is to quantify the distance between two vectors $\textbf{v}$ and $\textbf{w}$ using \textit{cosine distance} as follows:
\begin{align}
\label{eq:cosine}
d(\mathbf{v},\mathbf{w})= 1 - \frac{\mathbf{v}^T\mathbf{w}}{|\mathbf{v}||\mathbf{w}|}=1 - \frac{\mathbf{v}^T\mathbf{w}}{\sqrt{(\mathbf{v}^T\mathbf{v})(\mathbf{w}^T\mathbf{w})}},
\end{align} 
where $(.)^T$ is the transpose operation and $\mathbf{v}^T\mathbf{w}=\langle \mathbf{v},\mathbf{w} \rangle$ is the inner product of the two vectors. We first define the following vectors
\ifx \ncols \coltwo
\begin{align}
\label{eq:v_k}
&\mathbf{v}_k^\mathcal{N}(i)=\mathbf{x}_k(i)-\mathbf{c}_k^\mathcal{N} = \mathbf{x}_k(i)- {\sum_{\mathbf{x} \in \mathcal{N}_i^k} \mathbf{x}}/{|\mathcal{N}_k(i)|}, \\
\nonumber
&\mathbf{v}_k^\mathcal{X}(i)=\mathbf{c}^\mathcal{X}-\mathbf{x}_k(i) = \sum_{\mathbf{x} \in \mathcal{X}} \mathbf{x}/{|\mathcal{X}| - \mathbf{x}_k(i)},~\text{for }\mathcal{X} \in \{ \mathcal{S}, \mathcal{V}, \mathcal{F}\},\ 
\end{align}
\else
\begin{align}
\label{eq:v_k}
\nonumber
&\mathbf{v}_k^\mathcal{N}(i)=\mathbf{x}_k(i)-\mathbf{c}_k^\mathcal{N} = \mathbf{x}_k(i)- {\sum_{\mathbf{x} \in \mathcal{N}_i^k} \mathbf{x}}/{|\mathcal{N}_k(i)|}, \\
&\mathbf{v}_k^\mathcal{X}(i)=\mathbf{c}^\mathcal{X}-\mathbf{x}_k(i) = \sum_{\mathbf{x} \in \mathcal{X}} \mathbf{x}/{|\mathcal{X}| - \mathbf{x}_k(i)}, &\text{for }\mathcal{X} \in \{ \mathcal{S}, \mathcal{V}, \mathcal{F}\},\ 
\end{align}
\fi
where $\mathbf{v}_k^\mathcal{N}(i)$ is a vector pointing from the centroid of the current \textit{personalized normal cluster} ($\mathbf{c}_k^\mathcal{N}$) towards the current sample $\mathbf{x}_k(i)$ at time $k$ for patient $i$. Likewise, $\mathbf{v}_k^\mathcal{X}(i)$ is a vector from the current sample to $\mathbf{c}_\mathcal{X}$, the centroid of an abnormality class $\mathcal{X}$. A conceptual representation of these vectors in 2D space is depicted in Fig. \ref{fig:vecs}.

The relative tendency of a sample $\mathbf{x}_k(i)$ from normal cluster ($\mathcal{N}$) to any of the abnormal clusters $\mathcal{X} \in \{ \mathcal{S}, \mathcal{V}, \mathcal{F}\}$ is quantified by the cosine distance between $\mathbf{v}_k^\mathcal{N}(i)$, and any of the three vectors $\mathbf{v}_k^\mathcal{S}(i)$, $\mathbf{v}_k^\mathcal{V}(i)$, and $\mathbf{v}_k^\mathcal{F}(i)$.
More specifically, if a sample does not pass conditions in (\ref{eq:condition1}), then the \textit{personalized classifier} triggers a \textit{yellow} alarm as 
\begin{align}
\label{eq:personal_discrim}
\hat{y}^2_k(i) = \underset{\mathcal{X} \in \{ \mathcal{S}, \mathcal{V}, \mathcal{F} \}}{\text{argmin}}\{ d(\mathbf{v}_k(i),\mathbf{v}_{\mathcal{X}}(i)) \} 
\end{align}

Apparently, the relevance of cosine distance depends on the topology of the clusters in the feature space. This topology, itself is inherited from the feature extraction and feature selection methods. For example, as shown in Fig.\ref{fig:vecs}, the cosine distance suggests a highest alignment between the two vectors $\mathbf{v}_k^\mathcal{N}(i)$ and $\mathbf{v}_k^\mathcal{F}(i)$ in the original feature space, and hence $F$ is selected as the type of \textit{yellow} alarm, while apparently the sample leans more towards the abnormality cluster of type $V$. This inaccurate inference is due to the wider range of cluster $\mathcal{V}$ compared to $\mathcal{F}$. 
In order to eliminate the deviation analysis errors due to topological asymmetry, it is desired to transform the original topology into a more symmetric one, where cosine distances directly reflect the amount of deviations. 
\ifx \PSO \Yes
Here, we propose two methods to achieve the desired symmetry in the feature space.
\subsection{Numerically optimized nonlinear transformation} \label{sec:pso}

In this section, we propose a controlled non-linear spatial transformation, inspired by the \textit{kernel method}. 
\else
\subsection{Optimized nonlinear transformation} \label{sec:opt}
In this section, we propose an analytical method to optimize the non-linear transformation to achieve the desired clustering symmetry in the feature space.
\fi
The idea here is to map the feature vectors to a new space using a non-linear transformation. In contrast to the \textit{kernel method} used in SVM~\cite{shawe2004kernel}, we directly apply the non-linear function to feature vectors. Different functions can be used for this purpose, which may result in slightly different results. An exhaustive search for common kernels is a computationally expensive and unrealistic task. 
\ifx \PSO \Yes
A more efficient way to resolve this issue would be to search for an optimally weighted combination of a set of basis functions, such as polynomial and Gaussian functions~\cite{lanckriet2002robust}. This method has been proven to be robust and efficient if the coefficients are properly optimized~\cite{wang2014nonlinear}. 

In this work, polynomial function is arbitrarily used to implement and validate the proposed method. However, other non-linear functions can be used as well. The transformed feature vector $\mathbf{z}_k$ in the transformed space $\Omega^{d^\prime}$ of dimension $d^\prime$ is written as:

\begin{align}
\mathbf{z}_k
=\mathbf{\Psi_{w}} (\mathbf{x}_k) = 
\mathbf{w}^T \mathbf{\Psi} (\mathbf{x}_k)
=\sum_{l=1}^{d^\prime} w_l \psi_l (\mathbf{x}_k) \mathbf{\xi}_l,
\label{eq:z}
\end{align}
where $\mathbf{\Psi_{w}}:\Omega^d \mapsto \Omega^{d^\prime} $ is the utilized transformation, $\mathbf{w}=[w_1,\dots, w_{d^\prime}]$ is the normalized vector of coefficients with $|\mathbf{w}|=1$, $\psi_l()$ is the $l^{th}$ basis function and $\mathbf{\xi}_l=[\underbrace{0\dots 0}_{l-1}1 \underbrace{0\dots 0}_{d^\prime-l}]$ is the $l^{th}$ basis vector of $\Omega^{d^\prime}$. For instance, a kernel of type $k(\mathbf{x},\mathbf{y})=(1+\mathbf{x}^T\mathbf{y})^2$ in a 2-dimensional space with vectors $\mathbf{x}=[x_1~x_2]^T ,~\mathbf{y}=[y_1~y_2]^T \in \Omega^2$ can be represented as:
\begin{align}
\label{eq:example}
\nonumber
k(\mathbf{x},\mathbf{y})&=(1+\mathbf{x}^T\mathbf{y})^2= \langle \mathbf{\Psi_{w}} (\mathbf{x}),\mathbf{\Psi_{w}} (\mathbf{y}) \rangle\\
&=1+2x_1y_1+2x_2y_2+2x_1x_2y_1y_2+x_1^2y_1^2+x_2^2y_2^2
\end{align}
with 
\ifx \ncols \colone
\begin{align}
\nonumber
&\mathbf{\Psi_{w}} (\mathbf{x})=\mathbf{w}^T \mathbf{\Psi_{w}} (\mathbf{x}),~~~\mathbf{w}=[1~\sqrt{2}~\sqrt{2}~\sqrt{2}~1~1]^T,\\
&\phi_1(\mathbf{x})=1, \phi_2(\mathbf{x})=x_1,
\phi_3(\mathbf{x})=x_2,\phi_4(\mathbf{x})=x_1x_2,\phi_5(\mathbf{x})=x_1^2,\phi_6(\mathbf{x})=x_2^2.
\end{align}
\else
\begin{align}
\nonumber
&\mathbf{\Psi_{w}} (\mathbf{x})=\mathbf{w}^T \mathbf{\Psi_{w}} (\mathbf{x}),~~~\mathbf{w}=[1~\sqrt{2}~\sqrt{2}~\sqrt{2}~1~1]^T,\\
&\psi_1(\mathbf{x})=1, \psi_2(\mathbf{x})=x_1,
\psi_3(\mathbf{x})=x_2,\\
\nonumber
&\psi_4(\mathbf{x})=x_1x_2,\psi_5(\mathbf{x})=x_1^2,\psi_6(\mathbf{x})=x_2^2.
\end{align}

The desired properties of transformed data can be achieved by adjusting the coefficient vector $\mathbf{w}$. 

\fi
\else
Therefore, we choose a more straightforward based on the inverse of truncated \textit{logit} function as presented in section \ref{sec:mapping}.

In particular, we desire to develop a controlled transformation $T:\Omega^d \mapsto \Omega^{d^\prime}$ in order to map the original clusters into new ones to achieve the desired clustering properties. 

\fi

\ifx \ncols \coltwo
\begin{figure}
\centering
\includegraphics[width = 0.6\columnwidth]{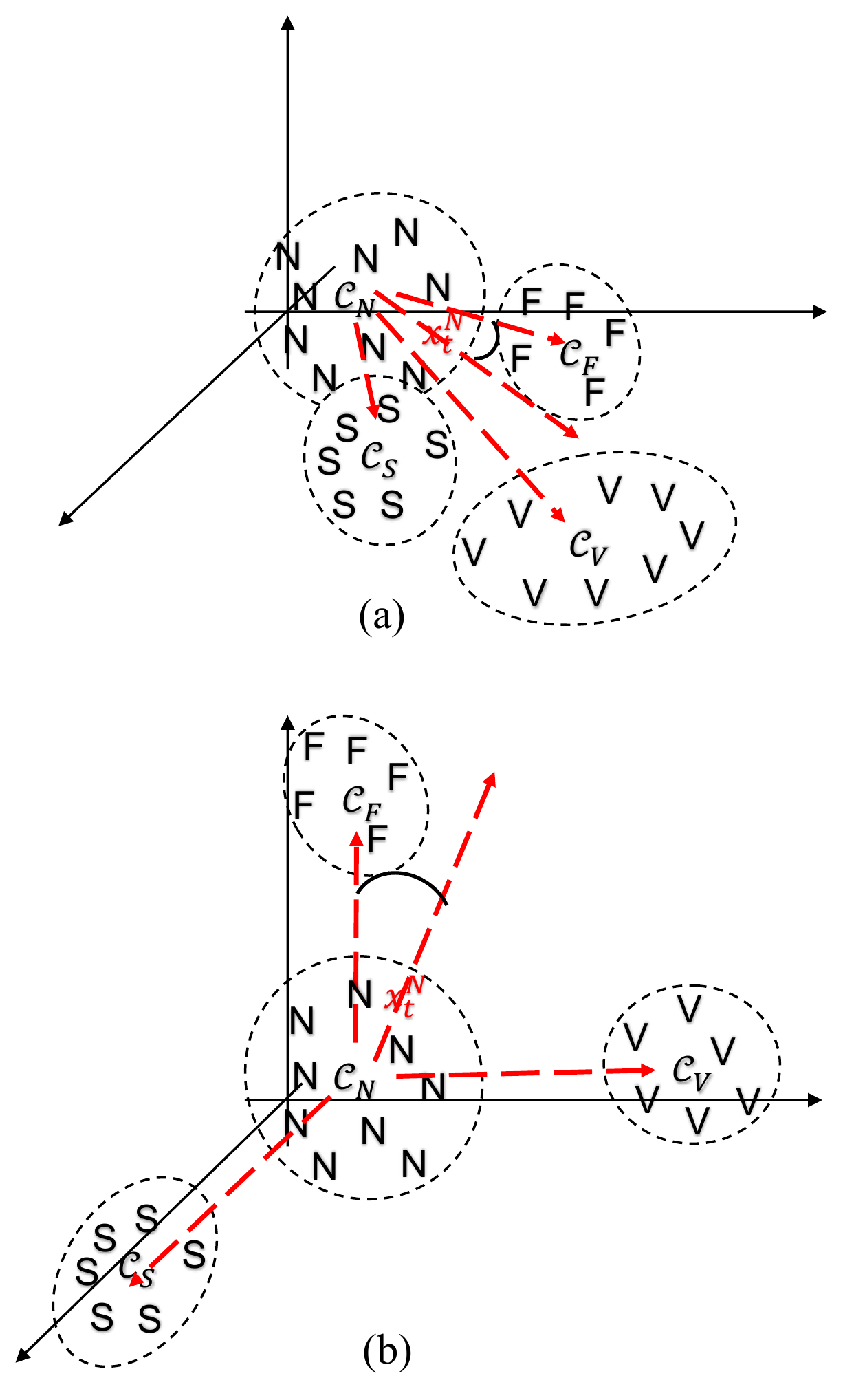}
\caption{Illustration of the clustering topology in (a) the original feature space, 
(b) the transformed feature space with the optimized mapping function. 
}
\label{fig:topo}
\end{figure}
\else
\begin{figure}
\centering
\includegraphics[width = \columnwidth]{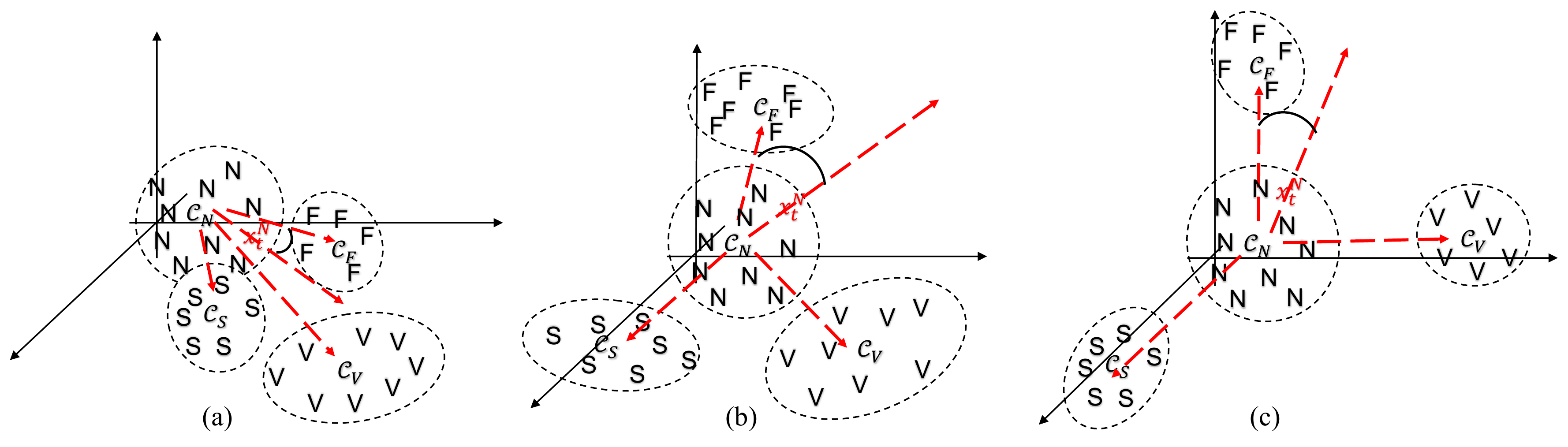}
\caption{Illustration of the cluster topology in (a) the original feature space, (b) the transformed feature space without reducing the within-cluster variance, and (c) the transformed feature space with the optimized mapping function. 
}
\label{fig:topo}
\end{figure}
\fi

\ifx \PSO \Yes
As illustrated in Fig. \ref{fig:topo}, the two desired clustering properties include the \textit{symmetry} and \textit{separability} of the clusters in the target space, which respectively are quantified with $o_1(\mathbf{w})$ and $o_2(\mathbf{w})$ as follows:

\ifx \ncols \colone
\begin{align}
\label{eq:obj}
\nonumber
&o_1(\mathbf{w}) = \frac{1}{\underset{c,d=2,\dots,C \text{ and } c\neq d }{\min}\{d(\mathbf{v}^{\mathcal{X}_c},\mathbf{v}^{\mathcal{X}_d})\}} &\text{(symmetry),} \\
&o_2(\mathbf{w}) = \frac{SW}{SB}=\frac{\sum_{c=1}^{C}  \sum_{\mathbf{z} \in \mathcal{X}_c}   (\mathbf{z}-\mathbf{c}^{\mathcal{X}_c})^T(\mathbf{z}-\mathbf{c}^{\mathcal{X}_c})}{\sum_{c=1}^{C}\sum_{d=1, d\neq c}^{C}  (\mathbf{c}^{\mathcal{X}_c}-\mathbf{c}^{\mathcal{X}_d})^T(\mathbf{c}^{\mathcal{X}_c}-\mathbf{c}^{\mathcal{X}_d}) } &\text{(separability),}
\end{align}
\else
\begin{align}
\label{eq:obj}
&o_1(T) = \frac{1}{\underset{c,d=2,\dots,C \text{ and } c\neq d }{\min}\{d(\mathbf{v}^{\mathcal{X}_c},\mathbf{v}^{\mathcal{X}_d})\}}\\
\nonumber
&o_2(T) = \frac{SW}{SB}=\frac{\sum_{c=1}^{C}  \sum_{\mathbf{z} \in \mathcal{X}_c}   (\mathbf{z}-\mathbf{c}^{\mathcal{X}_c})^T(\mathbf{z}-\mathbf{c}^{\mathcal{X}_c})}{\sum_{c=1}^{C}\sum_{d=1, d\neq c}^{C}  (\mathbf{c}^{\mathcal{X}_c}-\mathbf{c}^{\mathcal{X}_d})^T(\mathbf{c}^{\mathcal{X}_c}-\mathbf{c}^{\mathcal{X}_d}) }
\end{align}
\fi
where $\mathbf{v}^{\mathcal{X}_{c}}$ is a vector connecting the centroid of normal cluster to the abnormality clusters of type $c \in \{2,\dots,C\}$, defined as:
\begin{align}
\mathbf{v}^{\mathcal{X}_i} = \mathbf{c}_k^\mathcal{N} -  \mathbf{c}^{\mathcal{X}_i}
\end{align}

\else
As illustrated in Fig. \ref{fig:topo}, the two desired properties include the \textit{symmetry} and \textit{separability} of the clusters in the target space, which respectively are quantified with $o_1(T)$ and $o_2(T)$ as follows:

\ifx \ncols \colone
\begin{align}
\label{eq:obj}
\nonumber
&o_1(T) = \frac{1}{\underset{c,d=2,\dots,C \text{ and } c\neq d }{\min}\{d(\mathbf{v}^{\mathcal{X}_c},\mathbf{v}^{\mathcal{X}_d})\}} &\text{(symmetry),} \\
&o_2(T) = \frac{SW}{SB}=\frac{\sum_{c=1}^{C}  \sum_{\mathbf{z} \in \mathcal{X}_c}   (\mathbf{z}-\mathbf{c}^{\mathcal{X}_c})^T(\mathbf{z}-\mathbf{c}^{\mathcal{X}_c})}{\sum_{c=1}^{C}\sum_{d=1, d\neq c}^{C}  (\mathbf{c}^{\mathcal{X}_c}-\mathbf{c}^{\mathcal{X}_d})^T(\mathbf{c}^{\mathcal{X}_c}-\mathbf{c}^{\mathcal{X}_d}) } &\text{(separability),}
\end{align}
\else
\begin{align}
\label{eq:obj}
&o_1(T) = \frac{1}{\underset{c,d=2,\dots,C \text{ and } c\neq d }{\min}\{d(\mathbf{v}^{\mathcal{X}_c},\mathbf{v}^{\mathcal{X}_d})\}}\\
\nonumber
&o_2(T) = \frac{SW}{SB}=\frac{\sum_{c=1}^{C}  \sum_{\mathbf{z} \in \mathcal{X}_c}   (\mathbf{z}-\mathbf{c}^{\mathcal{X}_c})^T(\mathbf{z}-\mathbf{c}^{\mathcal{X}_c})}{\sum_{c=1}^{C}\sum_{d=1, d\neq c}^{C}  (\mathbf{c}^{\mathcal{X}_c}-\mathbf{c}^{\mathcal{X}_d})^T(\mathbf{c}^{\mathcal{X}_c}-\mathbf{c}^{\mathcal{X}_d}) }
\end{align}
\fi
where $\mathbf{z}_k=T(\mathbf{x}_k)$ is the transformed data sample, and $\mathbf{v}^{\mathcal{X}_{c}}$ is a vector connecting the centroid of normal cluster to the abnormality clusters of type $c \in \{2,\dots,C\}$ after the transformation, defined as:
\begin{align}
\mathbf{v}^{\mathcal{X}_i} = \mathbf{c}_k^\mathcal{N} -  \mathbf{c}^{\mathcal{X}_i}
\end{align}
\fi

Minimizing 
\ifx \PSO \Yes
$o_1(\mathbf{w})$ 
\else
$o_1(T)$ 
\fi
in (\ref{eq:obj}) maximizes the cosine distance between the vectors connecting the centroid of normal cluster to any of the abnormality clusters, which is achieved when the abnormality clusters \textit{symmetrically} surround the normal cluster. 
Likewise, 
\ifx \PSO \Yes
$o_2(\mathbf{w})$ 
\else
$o_2(T)$ 
\fi
is the inverse of \textit{Fisher discriminant} which enforces the \textit{separability} of abnormal clusters in terms of the ratio of within-cluster to between-cluster variances. Minimizing 
\ifx \PSO \Yes
$o_1(\mathbf{w})$ and $o_2(\mathbf{w})$ 
\else
$o_1(T)$ and $o_2(T)$ 
\fi
jointly improves the performance of the \textit{personalized classifier} and prevents the abnormal clusters from collapsing into one cluster.  
\ifx \PSO \Yes

This is a multi-objective optimization problem, which does not admit a closed form solution. For instance, $o_1(\mathbf{w})$ does not pass convexity test. On the other hand, the complexity of exhaustive search grows exponentially with 
$d^{\prime}$ and is not 
practically 
feasible for a large $d^{\prime}$. 

Here, we propose to use \textit{multi objective particle swarm optimization} (MOPSO) as a heuristic search algorithm to solve this problem~\cite{coello2002mopso}. 

We note that in PSO, each particle represents a randomly initialized vector of parameters. A global optima is achieved by iteratively moving each particle in a direction that is the linear combination of i) its local optima based on previous iterations, ii) the current global optima, and iii) the particle's previous direction. MOPSO extends this algorithm to a multi-objective setup. 

Here, each particle $P_j$ represents a coefficient vector $\mathbf{w}_j$ and we use $o_1(\mathbf{w})$ and $o_2(\mathbf{w})$ as objective functions to be minimized. At the end of each iteration $t$, we apply $\mathbf{w}_i^{(t)}\rightarrow \mathbf{w}_i^{(t)}/|\mathbf{w}_i^{(t)}|$ to impose the $|\mathbf{w}| =1$ condition. It is noteworthy that in contrast to classical PSO, where all particles converge to a unique global optima
, in MOPSO the particles converge to a \textit{Pareto front}, 
where $\beta o_1(\mathbf{w})+(1-\beta) o_2(\mathbf{w})$ is optimized with $0 \leq \beta\leq 1$.

In order to investigate the advantage of the proposed non-linear transformation, we applied the above-mentioned optimization method to optimize the coefficient vectors once for the original feature vectors $\mathbf{w}^T \mathbf{x}_k$ in $\Omega^d$ and again to the transformed feature vectors $\mathbf{\Psi_{w}} (\mathbf{x}_k)$ in $\Omega^{d^\prime}$. The former can be viewed as a special case of linear transformation. 
The results are shown in Fig. \ref{fig:pareto_compare}. Apparently, the resulting \textit{Pareto front} of the nonlinear model dominates that of the linear model with a significant margin. This improvement is due to the higher flexibility of the non-linear transformation to reshape the clustering geometry to \textit{separability} and \textit{symmetry}. Therefore, it improves the predictive capacity of the whole system, as confirmed by numerical results in section \ref{sec:result}.

\begin{figure}[t]
\centering
\includegraphics[scale=.55]{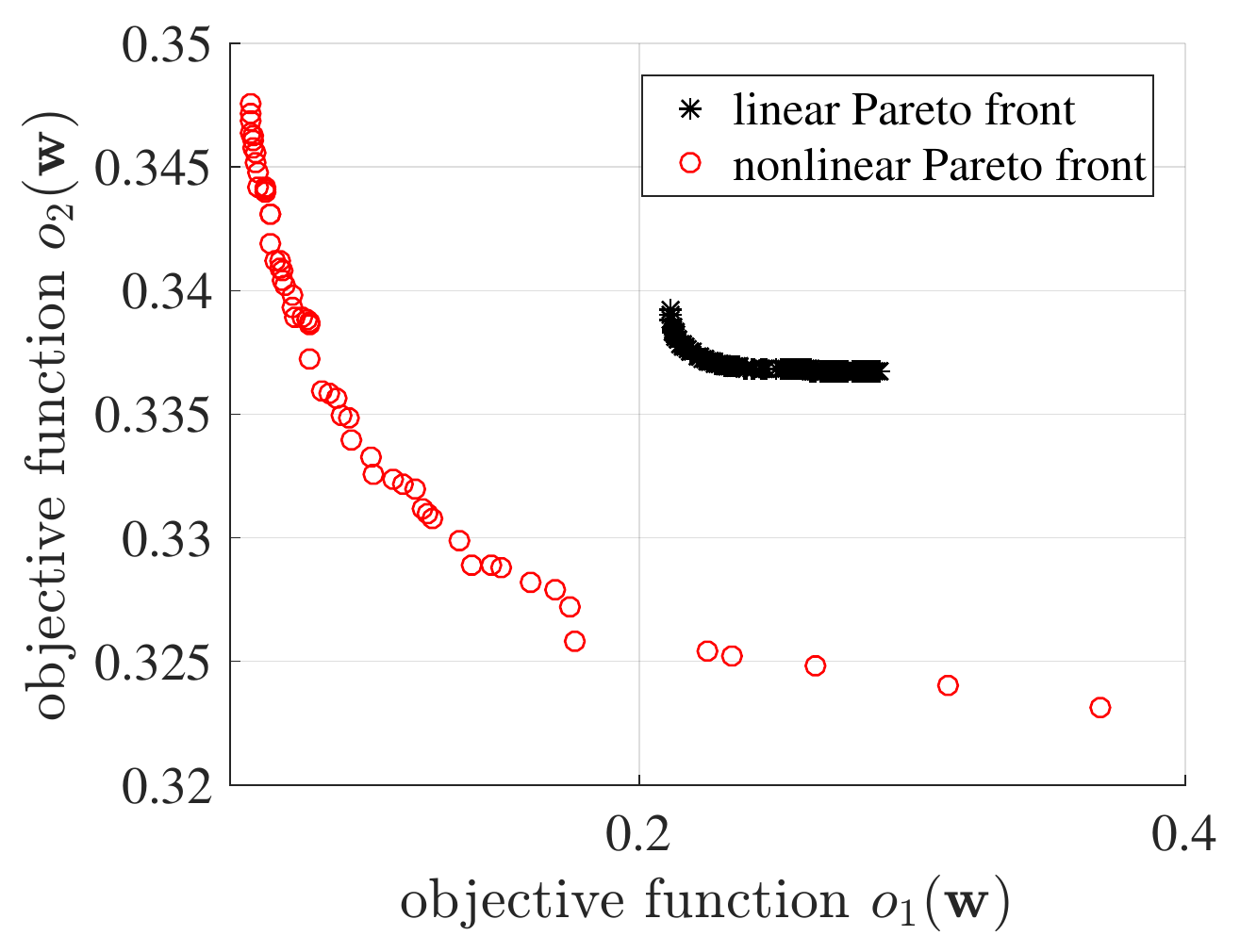}
\caption{Comparison between the \textit{Pareto front} of the MOPSO applied to linear and non-linear transformation (using polynomial functions).}
\label{fig:pareto_compare}
\end{figure}

\subsection{Optimized Deterministic Mapping Function}\label{sec:mapping}

The spatial transformation method presented in section \ref{sec:pso} uses computational optimization to achieve the desired geometry properties. In addition to the general drawbacks of heuristic methods (e.g. computational complexity, convergence issues, and the difficulty of interpreting the results), it is not straightforward to choose an appropriate set of basis function as the core of spatial mapping function. 
\else

This is a multi-objective optimization problem, which does not admit a closed form solution. In~\cite{chen2018predictive}, we developed a nonlinear transformation in terms of a linear combination of basis functions. We proposed a computational optimization method based on Multiple objective particle swarm optimization (MOPSO)~\cite{coello2002mopso} to optimize the coefficient vector $\mathbf{w}$ to achieve the desired symmetry. 

In addition to the general drawbacks of heuristic methods (e.g. computational complexity, convergence issues, and the difficulty of interpreting the results), it is not straightforward to choose an appropriate set of basis function as the core of spatial mapping function. 

\subsection{Design of Deterministic Mapping Function}\label{sec:mapping}
\fi

In this section, we propose a deterministic method that directly manipulates feature vectors in a tractable way to achieve a desired clustering geometry.

For the sake of simplicity, we use the \textit{hyper-spherical coordinate system}, where a $d$-dimensional position vector is represented by its radial distance from the origin as well as $d-1$ angles~\cite{nsphere}.  
More specifically, the vector $\mathbf{x}=[x_1~x_2~\cdots~x_d]^T$ in the \textit{Cartesian coordinate system} is represented as $\mathbf{x}=[r~\theta_1,\theta_2,\dots,\theta_{d-1}]^T$ in the \textit{hyper-spherical coordinate system} with the following direct and reverse relations in (\ref{eq:sph2cart}) and (\ref{eq:cart2sph}) \textcolor{black}{~\cite{nsphere}}: 
\begin{align}
\label{eq:sph2cart}
\begin{cases}
x_i = r \cos \theta_i \prod_{j=1}^{i-1}\sin \theta_j ,~~~~\text{for }i=1,2,\cdots,d-2,\\
x_d = r \sin \theta_{d-1} \prod_{j=1}^{d-2}\sin \theta_j,
\end{cases}
\end{align}

\begin{align}
\label{eq:cart2sph}
\begin{cases}
&r= \sqrt{\sum_{j=1}^d x_j^2}, \\
&\theta_i = \arccos \frac{x_i}{\sqrt{\sum_{j=i}^d x_j^2}},~~~~\text{for }i=1,2,\cdots,d-2,\\
&\theta_{d-1} = \text{sign}(x_d) \arccos \frac{x_{d-1}}{\sqrt{{x_d}^2+{x_{d-1}}^2}},
\end{cases}
\end{align}
where we have $0\leq \theta_j\leq\pi$ for $j=1,\dots ,d-2$; $0\leq \theta_{d-1}\leq 2 \pi$; and $0\leq r<\infty$.

The goal of the deterministic transformation $T:\Omega^d \mapsto \Omega^d$ is to map the original clusters into new ones such that the normal cluster resides in the origin 
(i.e. $\mathbf{c}^{\mathcal{N}}=\mathbf{0}$), and each abnormality cluster aligns with one of the coordinate system axes. The \textit{orthogonality} of abnormality clusters implies that the separation between the centroids of two clusters is  $1$ in terms of cosine distance, i.e. $d(\mathbf{c}^{\mathcal{X}_c},\mathbf{c}^{\mathcal{X}_d})=1$ for $c,d=2,3\dots C,~~c\neq d$. This property is achievable only if the number of abnormality clusters is less than the dimension of the feature space, i.e. ($C-1 \leq d$). This condition typically holds for most scenarios; here we have $C-1=3 \leq d=8$. 
The second desired property is the \textit{concentration} of the abnormality clusters around the corresponding coordinate axis (i.e. basis vectors of Euclidean space) to achieve a minimal overlap between adjacent clusters. These properties are illustrated in Fig. \ref{fig:topo}.

Each cluster in the feature space is represented by its centroid locations, namely $\mathbf{c}_k^\mathcal{N}(i),~ \mathbf{c}^{\mathcal{V}},~ \mathbf{c}^{\mathcal{S}},~ \mathbf{c}^{\mathcal{F}}$, respectively for the normal cluster at time $k$ for user $i$ and the three abnormality clusters of type $V$, $S$, and $F$. Hereafter, we omit the 
user index $i$ for notation convenience, when it is clear from the context. 

Firstly, we shift all clusters such that the centroid of normal cluster coincides with the origin. Obviously this linear transformation does not change the clustering geometry and the subsequent deviation analysis. The clustering topology in the original feature space can be equivalently represented by the following matrix with three column vectors:
\begin{align}
C = [\mathbf{c}_{\mathcal{V}} - \mathbf{c}_N^k ~\mathbf{c}_{\mathcal{S}} - \mathbf{c}_N^k ~\mathbf{c}_{\mathcal{F}} - \mathbf{c}_N^k]= 
\left[\mathbf{v}_{\mathcal{VN}}~
\mathbf{v}_{\mathcal{SN}} ~
\mathbf{v}_{\mathcal{FN}} \right]
\end{align}

In order to impose the \textit{orthogonality} property and ensure a full symmetry, we desire to develop a transformation that maps the columns vectors of $C$ into a set of orthogonal vectors. Therefore, we can use the popular method of \textit{Gram-Schmidt} orthogonalization method explained in~\cite{Gram-schmidth2}
\begin{align}
C^{\perp}= \text{Gram-Schmidt}(C)
=\left[\mathbf{v}_{\mathcal{VN}}^{\perp}~\mathbf{v}_{\mathcal{SN}}^{\perp}~\mathbf{v}_{\mathcal{FN}}^{\perp}\right]
\end{align}

Now, the goal here is to develop a non-linear transform $T:\Omega^d \mapsto \Omega^d$, which maps the columns of $C$ to $C^{\perp}$, namely
\begin{align}
T(\mathbf{v}_{\mathcal{VN}})=\mathbf{v}_{\mathcal{VN}}^{\perp},~T(\mathbf{v}_{\mathcal{SN}})=\mathbf{v}_{\mathcal{SN}}^{\perp},~T(\mathbf{v}_{\mathcal{FN}})=\mathbf{v}_{\mathcal{VN}}^{\perp},
\end{align}
while improving the \textit{concentration} property. Since the direction of the first vector remains unchanged under the \textit{Gram-Schmidt} algorithm (i.e. $\mathbf{v}_{\mathcal{VN}} = \mathbf{v}_{\mathcal{VN}}^{\perp}$), it is sufficient to satisfy the following equations:
\begin{align}
\label{eq:constraints}
\begin{aligned}
T(\mathbf{v}_{\mathcal{SN}} - \mathbf{v}_{\mathcal{VN}}) &= \mathbf{v}_{\mathcal{SN}}^{\perp} - \mathbf{v}_{\mathcal{VN}}^{\perp} &= \mathbf{v}_{\mathcal{SN}}^{\perp} - \mathbf{v}_{\mathcal{VN}}\\
T(\mathbf{v}_{\mathcal{FN}} - \mathbf{v}_{\mathcal{VN}}) &= \mathbf{v}_{\mathcal{FN}}^{\perp} - \mathbf{v}_{\mathcal{VN}}^{\perp} &= \mathbf{v}_{\mathcal{FN}}^{\perp} - \mathbf{v}_{\mathcal{VN}}
\end{aligned}
\end{align}

In order to simplify the process, it is more convenient to work with the angular coordinates, therefore we represent the matrices $C$ and $C^{\perp}$ in terms of their hyper-spherical coordinate system using (\ref{eq:cart2sph}) as follows:
\begin{align}
C^{\perp}_{*} =
\begin{bmatrix}
    r_{\mathcal{VN}}^{\perp} &  r_{\mathcal{SN}}^{\perp} &   r_{\mathcal{FN}}^{\perp} \\
    \theta_{1_{\mathcal{VN}}}^{\perp} & \theta_{1_{\mathcal{SN}}}^{\perp} & \theta_{1_{\mathcal{FN}}}^{\perp} \\
      \vdots & \vdots & \vdots  \\
    \theta_{{d-1}_{\mathcal{VN}}}^{\perp} & \theta_{{d-1}_{\mathcal{SN}}}^{\perp} & \theta_{{d-1}_{\mathcal{FN}}}^{\perp} \\
\end{bmatrix}
\end{align}

Furthermore, since the \textit{orthogonality} of vectors is independent of their radii $r$, we can decompose the function $T$ into $d-1$ subfunctions $T_i,~i=1,2,\dots, d-1$, each of which applies to one of the $d-1$ angular dimensions ($\theta_1,\dots,\theta_{d-1}$) and satisfy (\ref{eq:constraints}). Finally, as a separate step, we can develop a function $T_r$ to achieve the desired radius symmetry, $r_{\mathcal{VN}}^{\perp}=r_{\mathcal{SN}}^{\perp}=r_{\mathcal{FN}}^{\perp}=1$. 

\begin{figure}[thpb]
\centering
\ifx \ncols \colone
\includegraphics[scale=0.6]{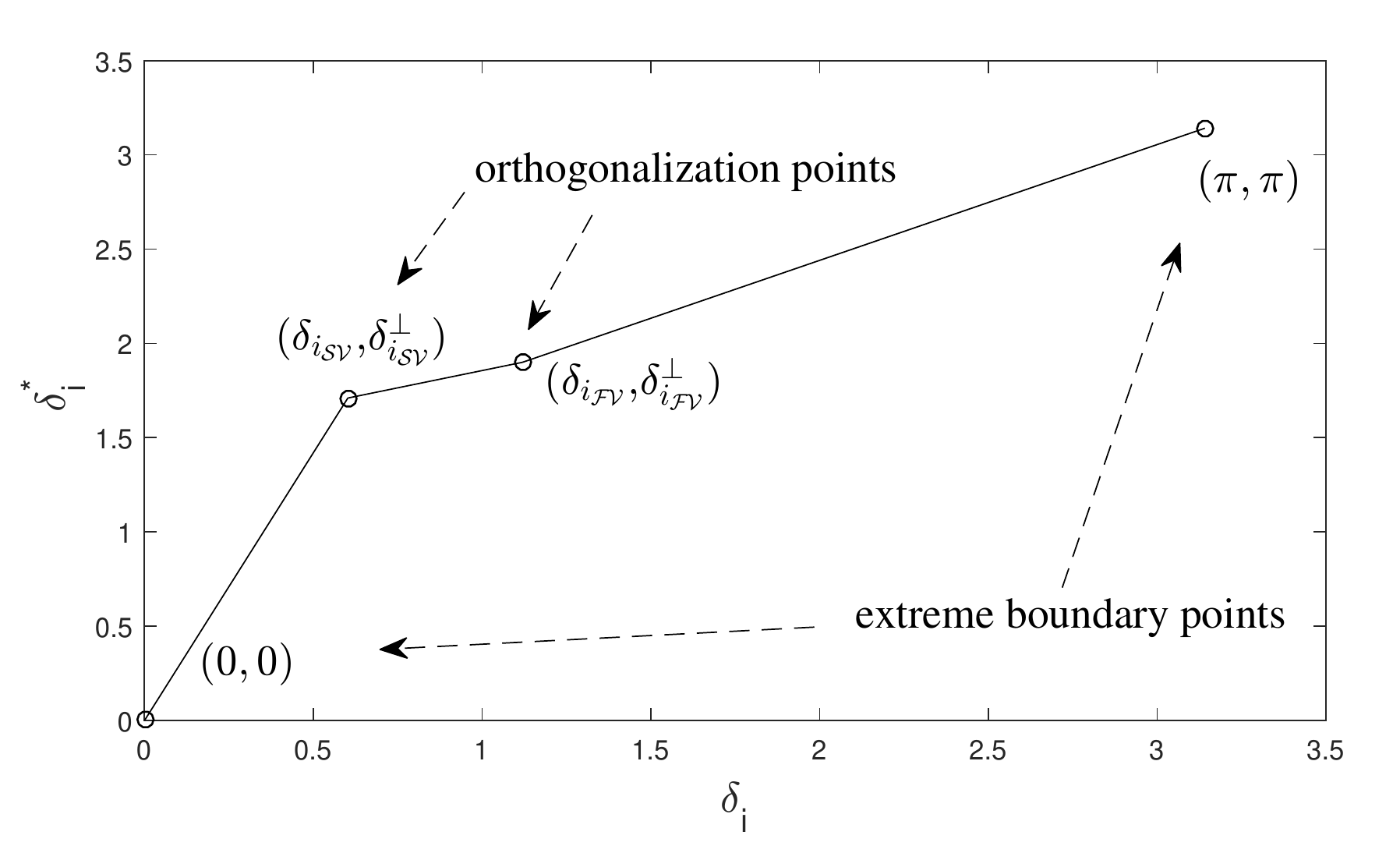}
\else
\includegraphics[scale=0.42]{figures/simple_spline2.pdf}
\fi
\caption{The simple mapping function for one angular dimension which maps the \textit{target points} in the original space 
to the desired \textit{target points} using piece-wise linear functions.}
\label{fig:simple_spline}
\end{figure}

We use the notation $\mathbf{v}_{\mathcal{SN}} - \mathbf{v}_{\mathcal{VN}} = \Delta_{\mathcal{SV}}$ and denote the the $i^{th}$ angular dimension of $\Delta_{\mathcal{SV}}$ as $\delta_{i_{{\mathcal{SV}}}}$. We use similar notations for the other dimensions, (e.g. $\mathbf{v}_{\mathcal{FN}} - \mathbf{v}_{\mathcal{VN}}$). 
For each angular dimension $i$, $T_i$ passes through points ($\delta_{i_{{\mathcal{SV}}}}$,$\delta^{\perp}_{i_{{\mathcal{SV}}}}$) and ($\delta_{i_{{\mathcal{FV}}}}$,$\delta^{\perp}_{i_{{\mathcal{FV}}}}$), as well as the two \textit{extreme boundary points} defining the domain of the function, which include $(0,0)$ for the lower limit and $(\pi,\pi)$ and $(2\pi,2\pi)$ for the upper limit, respectively for $=1,2\dots,d-2$ and $i=d-1$. We call the collection of these points as \textit{target points}, as shown in Fig. \ref{fig:simple_spline}.

\noindent\textbf{Piece-wise linear mapping:} In order to preserve a maximal similarity between the original and the transformed clustering geometry, it is preferred to use continuous, monotonic and smooth functions. 
One natural option is to use piece-wise linear function, which passes through the {target points}, as depicted in Fig. \ref{fig:simple_spline}. 

\noindent\textbf{Optimized Mapping Function:} 
The piecewise linear mapping function $T_i$ in Fig. \ref{fig:simple_spline} exhibits the two desired properties of \textit{monotonicity} and \textit{continuity}. However, it suffers from some drawbacks. Firstly, the function is not differentiable at the \textit{orthogonalization} points ($\delta_{i_{{\mathcal{SV}}}}$,$\delta^{\perp}_{i_{{\mathcal{SV}}}}$) and ($\delta_{i_{{\mathcal{FV}}}}$,$\delta^{\perp}_{i_{{\mathcal{FV}}}}$), which can lead to severe cluster deformation (and even mapping convex-clusters into non-convex clusters). Secondly, to provide maximal separation among transformed clusters, it is unavoidable to include a level of \textit{nonlinearity} that concentrates the angular range of an original cluster into the proximity of the target angle of the corresponding \textit{orthogonalization} point. 
However, there is a trade-off between the level of \textit{concentration} and the \textit{linearity}, which needs to be carefully addressed. 
A proper mapping function should satisfy 
the following mathematical conditions:

\begin{itemize}
\item the function is continuous and monotonic increasing 
\item the function passes through all \textit{target points};
\item the function is differentiable at target points;
\item the derivative of the function is small at the target points, which correspond to the centroids of clusters in the original and transformed space (i.e. $\delta_{i_{{\mathcal{SV}}}}$ and $\delta_{i_{{\mathcal{FV}}}}$); 
\item the derivative of the function is large at the boundaries of two regions (points $(\epsilon_{i_{{\mathcal{SV}}}},\epsilon^{\perp}_{i_{{\mathcal{SV}}}})$ and $(\gamma_{i_{{\mathcal{SV}}}},\gamma^{\perp}_{i_{{\mathcal{SV}}}})$. 
\end{itemize}

\begin{figure}[thpb]
\centering
\ifx \ncols \colone
\includegraphics[scale=.6]{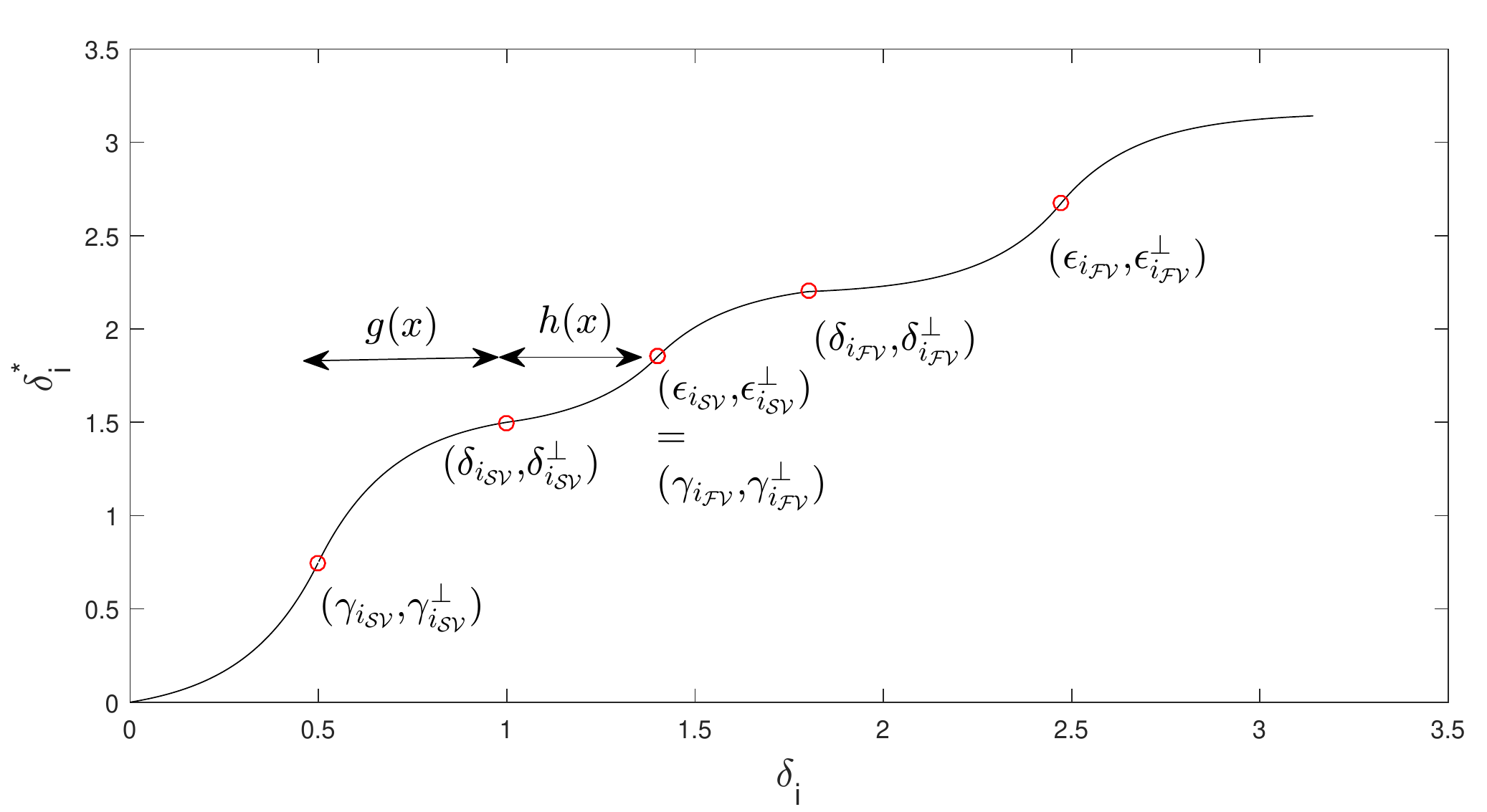}
\else
\includegraphics[scale=.4]{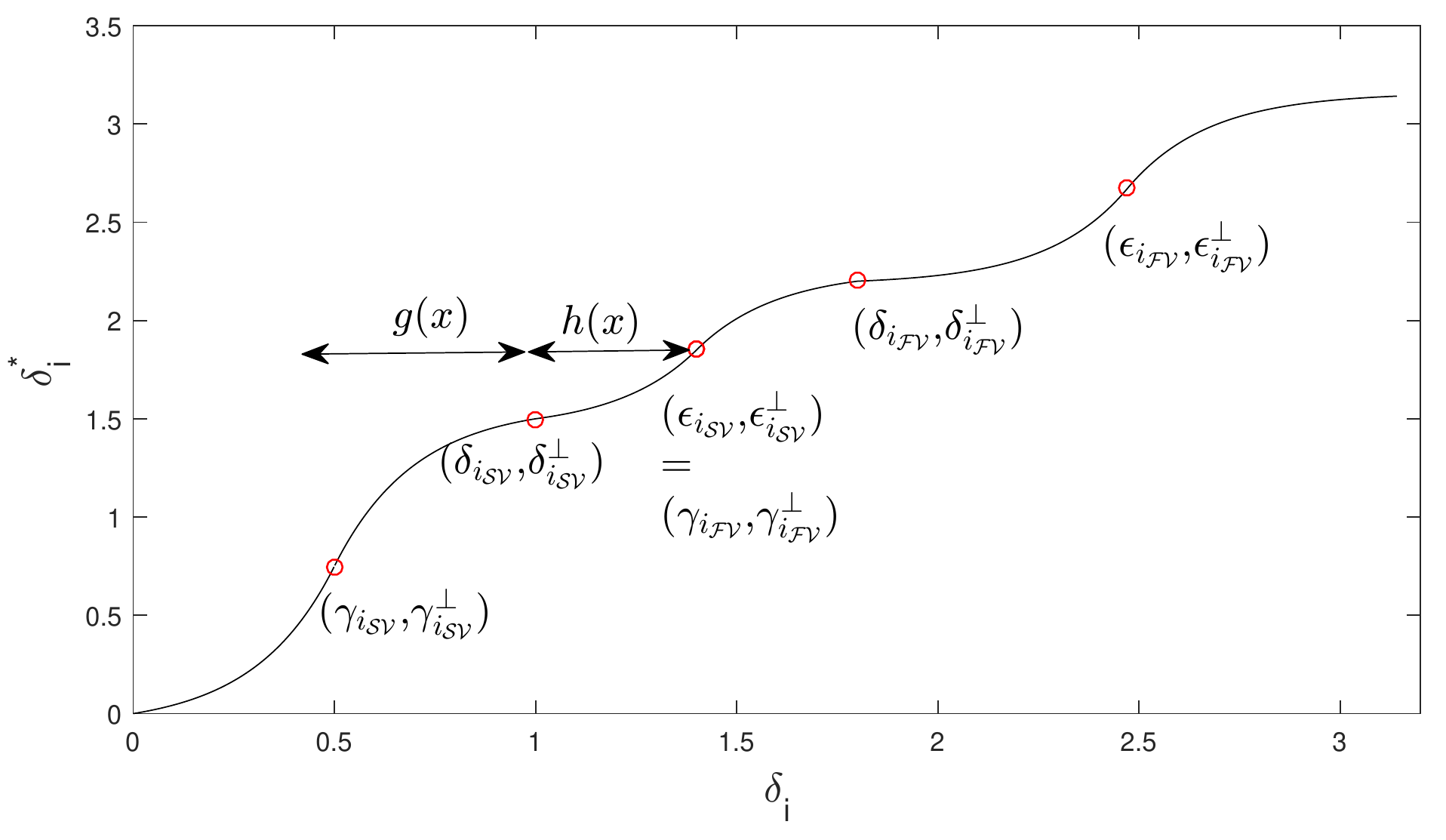}
\fi
\caption{Optimized Mapping Function $f$}
\label{fig:optimized_map}
\end{figure}

We split the domain of the function, i.e. $[0,\pi]$ for $i=1,2,\dots,d-2$ and $[0,2\pi]$ for $i=d-1$ into regions $L_j=[\gamma_j,\epsilon_j]$ around the \textit{orthogonalization} points $\delta_j$. The \textit{lower boundary point} $\gamma_j$ is simply determined as the midpoint between the current and the previous \textit{target point}, i.e. $\gamma_j=\frac{1}{2}(\delta_j+\delta_{j-1})$. Likewise, \textit{upper boundary point} $\epsilon_j$ is defined as $\epsilon_j=\frac{1}{2}(\delta_j+\delta_{j+1})$. Obviously, we have $\epsilon_j=\gamma_{j+1}$. These points are depicted in Fig. \ref{fig:optimized_map} and for our scenario include:
\begin{align}
\label{eq:midpoint}
(\epsilon_{i_{{\mathcal{SV}}}},\epsilon^{\perp}_{i_{{\mathcal{SV}}}}) =
(\gamma_{i_{{\mathcal{FV}}}},\gamma^{\perp}_{i_{{\mathcal{FV}}}}) =
(\frac{\delta_{i_{{\mathcal{SV}}}} + \delta_{i_{{\mathcal{FV}}}}}{2}, \frac{\delta^{\perp}_{i_{{\mathcal{SV}}}} + \delta^{\perp}_{i_{{\mathcal{FV}}}}}{2}).
\end{align}

For each region $L_j$, we define two functions $h_j(x)$ and $p_j(x)$, respectively, for intervals $[\gamma_j,\delta_j]$, and $[\delta_j,\epsilon_j]$.
We choose to use the inverse of \textit{logit} function defined as follows (after omitting the subscript $j$):
\begin{align}\label{eq:basicfunctions}
\nonumber
g(x;\alpha_g,\gamma,\gamma^{\perp},\delta,\delta^{\perp}) &= K_g[e^{\alpha_g(\delta-x,0)^{+}}-1] + \delta^{\perp} \\
h(x;\alpha_h,\epsilon,\epsilon^{\perp},\delta,\delta^{\perp}) &= K_h[e^{\alpha_h(x-\delta,0)^{+}}-1] + \delta^{\perp}
\end{align}
with 
\begin{align}\label{eq:basicfunctions-params}
\nonumber
K_g &= \frac{\gamma^{\perp}-\delta^{\perp}}{e^{\alpha_g(\delta-\gamma,0)^{+}}-1}\\
K_h &= \frac{\epsilon^{\perp}-\delta^{\perp}}{e^{\alpha_h(\epsilon-\delta,0)^{+}}-1}
\end{align}

It is easy to verify that these functions pass through the target points, since we have $g(\delta_j)=h(\delta_j)=\delta_j^{\perp}$, $g(\gamma_j)=\gamma_j^{\perp}$, and $h(\epsilon_j)=\epsilon_j^{\perp}$. Parameters $\alpha_g$ and $\alpha_h$ control the sharpness of the functions $g()$ and $h()$ and should be tunned to balance between the \textit{concentration} and \textit{linearity} of the functions. To achieve linearity, we can set $\alpha \rightarrow 0$ noting that $e^{x}\approx 1+x$ for small $x$. Finally, to obtain differentiability at target point $\delta_j$, the right derivative of $g_j(x)$ should be equal to the left derivative of $h_j(x)$, which is achieved by setting $\alpha_g K_g = \alpha_h K_h$. We set $\alpha_g=1$ in this work. 
Fig. \ref{fig:optimized_map} illustrates an implementation of this function for the entire range $[0,\pi]$.


\section{Numerical Results}
\label{sec:result}
In this section, the performance of the proposed method is evaluated in terms of two aspects. We first analyze the classification performance of the system and compare it against the state of the art classification results using the same ECG dataset.  
Secondly, the prediction power of the proposed method 
is evaluated.

\subsection{Classification Performance}
The original test dataset DS2 contains 15357 samples after feature extraction. While training the \textit{personalized classifier}, the normally labeled samples within the the first 5 minutes of the signal serve as initialization set for the \textit{personalized dynamic normal} cluster. After excluding the first 5 minute for all test signals, the resulting dataset includes 
$10105$ type-N, $1702$ type-V, $508$ type-S and $99$ type-F samples (a total of $12414$ samples).


Table \ref{table:classification_cumu} summarizes the obtained cumulated confusion matrix for all records in the test set using our propose method. Results are provided in terms of \textit{red alarms} as the output of the first stage (the \textit{global classifier}) as well as the final labels (after deviation analysis), where \textit{yellow} and \textit{red} alarms are combined. It is seen that part of the normally labeled samples by the \textit{global classifier} are assigned with \textit{yellow} alarms in the final results, which shows the role of the \textit{personalized classifier}.


\if \ncols \colone
\begin{table}[t]
\footnotesize
	\centering
	\caption{Cumulated Confusion Matrix for All Records in DS2. Results are provided in terms of \textit{red} alarms (by the \textit{global classifier}) as well as the final results.}
	\begin{tabular}{|l|l|c|c|c|c|l|l|c|c|c|c|}
		\hline 
		&  \multicolumn{5}{c|}{Ground Truth} & &\multicolumn{5}{c|}{Ground Truth} \\ 
        \hline 
        \multirow{5}{5em}{Results (\textit{Global Classifier})} &  & N & V & S & F  & \multirow{5}{5em}{Results (Final Labels)} & & N & V & S & F \\ 
        \cline{2-6}  \cline{8-12} 
		& N & 10076& 38 & 90 & 5 & & N & 9255& 21 & 72 & 1 \\ 
        \cline{2-6}  \cline{8-12} 
		& V & 22 & 1663 & 2 & 7 & & V & 657 & 1678 & 8 & 9  \\ 
        \cline{2-6}  \cline{8-12} 
		& S & 6 & 1 & 416  & 0 &  & S & 71 & 3 & 417 & 0  \\ 
        \cline{2-6}  \cline{8-12} 
        & F & 1 & 0 & 0 & 87 &  & F & 122 & 0 & 11 & 89  \\ 
        \cline{2-6}  \cline{8-12} 
        \hline        
	\end{tabular}
	\label{table:classification_cumu} 
\end{table}
\else
\begin{table}[t]
\footnotesize
	\centering
	\caption{Cumulated Confusion Matrix for All Records in DS2.}
	\begin{tabular}{|m{6em}|l|c|c|c|c|}
		\hline 
		&  \multicolumn{4}{c}{Ground Truth} &\\ 
        \hline
		\multirow{5}{6em}{Result (\textit{Global Classifier})} &  & N & V & S & F  \\\cline{2-6}
		& N & 10076& 38 & 90 & 5 \\\cline{2-6} 
		&V & 22 & 1663 & 2 & 7  \\\cline{2-6}
		&S & 6 & 1 & 416 & 0  \\\cline{2-6}
        &F& 1 & 0 & 0 & 87  \\\hline
        \hline
		\multirow{5}{6em}{Results (Final Labels)} &  & N & V & S & F  \\\cline{2-6}
		& N & 9255& 21 & 72 & 1 \\\cline{2-6} 
		&V & 657 & 1678 & 8 & 9  \\\cline{2-6}
		&S & 71 & 3 & 417 & 0  \\\cline{2-6}
        &F& 122 & 0 & 11 & 89  \\\hline        
	\end{tabular}
	\label{table:classification_cumu} 
	\vspace{-0.15in}
\end{table}
\fi

In order to measure the classification performance, we adopt three metrics: accuracy ($ACC=\frac{TP+TN}{TP+TN+FN+FP}$), sensitivity ($SE= \frac{TP}{TP+FN}$), and specificity ($SP=\frac{TN}{TN+FP}$)
, as proposed in~\cite{deChazal2006}. 
The metrics are calculated based on the true positive ($TP$), false positive ($FP$), false negative ($FN$) and true negative ($TN$) using a binary confusion matrix, where one class is the specific abnormality class and all other abnormality and normal classes combined into one class (i.e. after converting the 4x4 confusion matrix to a 2x2 matrix for each class).

Table \ref{table:variation} evaluates the performance of the proposed method for different metrics for each class. The results demonstrate the high classification accuracy of the proposed system in terms of different performance metrics. Furthermore, to evaluate the robustness of the proposed method, the performance variation over $22$ test records in DS2 is presented by providing the medians and interquartile range (IQRs) for each metric and each class. 
We observe that the proposed method demonstrates a stable performance for all abnormality classes. The stability is higher for class V. 
\if \ncols \colone 
\begin{table*}[t]
\footnotesize
\centering
\caption{Classification Performance and Within-Set Variation of Proposed System}
\label{table:variation}
\resizebox{\columnwidth}{!}{
\begin{tabular}{|c|c|c|c|c|c|c|c|c|c|c|c|c|}
\hline
\multirow{2}{*}{statistics} & \multicolumn{3}{c|}{N}                  & \multicolumn{3}{c|}{V}                  & \multicolumn{3}{c|}{S}                  & \multicolumn{3}{c|}{F}                  \\ \cline{2-13} 
                            & \textit{ACC} & \textit{SE} & \textit{SP} & \textit{ACC} & \textit{SE} & \textit{SP} & \textit{ACC} & \textit{SE} & \textit{SP} & \textit{ACC} & \textit{SE} & \textit{SP} \\ \hline
cumulated                   & 92.4        & 91.59       & 95.93       & 94.38       & 98.59       & 93.71       & 98.67       & 82.09       & 99.38       & 98.85       & 89.9        & 98.92       \\ \hline
median                      & 94.45       & 92.21       & 95.42       & 96.17       & 99.55       & 95.71       & 99.38       & 80.65       & 99.84       & 99.11       & 90.91       & 99.11       \\ \hline
IQR                         & 6.33        & 10.08       & 11.91       & 5.17        & 1.64        & 8.62        & 1.76        & 19.35       & 0.61        & 1.58        & 23.33       & 1.49        \\ \hline
\end{tabular}}
\end{table*}
\else
\begin{table*}[t]
\footnotesize
\centering
\caption{Classification Performance and Within-Set Variation of Proposed System}
\label{table:variation}
\begin{tabular}{|c|c|c|c|c|c|c|c|c|c|c|c|c|}
\hline
\multirow{2}{*}{statistics} & \multicolumn{3}{c|}{N}                  & \multicolumn{3}{c|}{V}                  & \multicolumn{3}{c|}{S}                  & \multicolumn{3}{c|}{F}                  \\ \cline{2-13} 
                            & \textit{ACC} & \textit{SE} & \textit{SP} & \textit{ACC} & \textit{SE} & \textit{SP} & \textit{ACC} & \textit{SE} & \textit{SP} & \textit{ACC} & \textit{SE} & \textit{SP} \\ \hline
cumulated                   & 92.4        & 91.59       & 95.93       & 94.38       & 98.59       & 93.71       & 98.67       & 82.09       & 99.38       & 98.85       & 89.9        & 98.92       \\ \hline
median                      & 94.45       & 92.21       & 95.42       & 96.17       & 99.55       & 95.71       & 99.38       & 80.65       & 99.84       & 99.11       & 90.91       & 99.11       \\ \hline
IQR                         & 6.33        & 10.08       & 11.91       & 5.17        & 1.64        & 8.62        & 1.76        & 19.35       & 0.61        & 1.58        & 23.33       & 1.49        \\ \hline
\end{tabular}
\end{table*}
\fi

To further evaluate the performance of the proposed method, we compare the classification results of our method against 5 significant classifiers applied to the ECG recording in the benchmark MITDB dataset. According to AAMI standards, the performance of ECG classification should be evaluated over the binary classifiers applied to \textit{Ventricular (V)} versus \textit{non-V}types and \textit{Supraventricular (S)} versus \textit{non-S} types. 
To enable a fair comparison, we select 11 ECG records which are common among all methods and compare the median of each classification metrics over these 11 records. Since our method includes two sequential classification stages, we 
use the final labels here. 
The comparison results are presented in Table \ref{table:classification_comp}. Overall, the proposed method shows a higher sensitivity for both types V and S. Especially for type S, the proposed method shows an advantage over all three metrics compared to the five reference methods. The results for Class V are also comparable to other methods.

\if \ncols \colone
\begin{table}[t]
\footnotesize
\centering
\caption{Classification performance compared with five algorithms in literature using 11 common records in MITDB.}
\label{table:classification_comp}
\begin{tabular}{|c|c|c|c|c|c|c|}
\hline
\multirow{2}{*}{Methods} & \multicolumn{3}{c|}{Class V} & \multicolumn{3}{c|}{Class S} \\ \cline{2-7} 
                         & ACC     & SE     & SP     & AC      & SE     & SP     \\ \hline
Proposed                 & 96.6   & 98.2   & 92.4   & 98.63   & 88.89  & 99.41  \\ \hline
Hu \textit{et al.}~\cite{Hu_et_al}     & 94.8   & 78.9   & 96.8   & N/A     & N/A    & N/A    \\ \hline
de Chazal \textit{et al.}~\cite{autofs}  & 96.4   & 77.5   & N/A    & N/A     & N/A    & N/A    \\ \hline
Jiang and Kong~\cite{bbnn}    & 98.8   & 78.9   & 96.8   & 97.5    & 74.9   & 98.8   \\ \hline
Ince \textit{et al.}~\cite{ince2009generic}    & 97.9   & 90.3   & 98.8   & 96.1    & 81.8   & 98.5   \\ \hline
Kiranyaz \textit{et al.}~\cite{Kiranyaz}         & 98.9   & 95.9   & 99.4   & 96.4    & 68.8   & 99.5   \\ \hline
\end{tabular}
\end{table}
\else
\begin{table*}[t]
\footnotesize
\centering
\caption{Classification performance compared with five algorithms in literature using 11 common records in MITDB.}
\label{table:classification_comp}
\begin{tabular}{|c|c|c|c|c|c|c|}
\hline
\multirow{2}{*}{Methods} & \multicolumn{3}{c|}{Class V} & \multicolumn{3}{c|}{Class S} \\ \cline{2-7} 
                         & ACC     & SE     & SP     & AC      & SE     & SP     \\ \hline
Proposed                 & 96.6   & 98.2   & 92.4   & 98.63   & 88.89  & 99.41  \\ \hline
Hu \textit{et al.}~\cite{Hu_et_al}     & 94.8   & 78.9   & 96.8   & N/A     & N/A    & N/A    \\ \hline
de Chazal \textit{et al.}~\cite{autofs}  & 96.4   & 77.5   & N/A    & N/A     & N/A    & N/A    \\ \hline
Jiang and Kong~\cite{bbnn}    & 98.8   & 78.9   & 96.8   & 97.5    & 74.9   & 98.8   \\ \hline
Ince \textit{et al.}~\cite{ince2009generic}    & 97.9   & 90.3   & 98.8   & 96.1    & 81.8   & 98.5   \\ \hline
Kiranyaz \textit{et al.}~\cite{Kiranyaz}         & 98.9   & 95.9   & 99.4   & 96.4    & 68.8   & 99.5   \\ \hline
\end{tabular}
\end{table*}
\fi

\subsection{Prediction Performance}
A unique feature of the proposed method is its predictive capacity.  
In order to test this capability, 
all beats following a \textit{yellow} alarm of a specific type is investigated to asses the chance of upcoming \textit{red} alarms. This process is repeated for \textit{yellow} alarms of all types. We only account for the first abnormality type, which occurs after the \textit{yellow} alarm. As we used confusion matrix to evaluate the classification accuracy, the performance of the prediction can be summarized by a confusion matrix for the 3 abnormality types. Probabilities of observing a certain type of an abnormal beat after a \textit{yellow} alarm is calculated using the predictive confusion matrix and compared to the prior probability of observing an abnormality of the same type. This process is formulated in the following two equations:

\begin{align}
\nonumber 
& P(\hat{y}_{k+i}=X_r|\hat{y}_k=X_y)=\frac{\text{\# of } y_{k+i}=X \text{ after } \hat{y}_k=X_y}{\text{\# of true alarms after }\hat{y}_k=X_y} \\
& P(\hat{y}_{k+i}=X_r)=\frac{\text{\# of true alarm of type } X~ (y_{k}=X)}{\text{\# of all true alarms } (y_{k}\neq N)} 
\end{align}

The prediction power of the proposed method for each abnormality type $X$ is evaluated by comparing $P(\hat{y}_{k+i}=X_r|\hat{y}_{k}=X_y)$ and $P(\hat{y}_{k+i}=X_r)$. A shown in Table \ref{table:pred}, the probability of observing a certain type of abnormality after a \textit{yellow} alarm of the same type is higher than its prior probability. For example, without knowing the type of a preceding \textit{yellow} alarm, the probability of observing a type $V$ sample is 71.54\%, while the probability of observing a type $V$ sample after observing a \textit{yellow} alarm of type $V$ is 77.45\% ($5.91\%$ higher than the prior probability). This improvement holds consistently among all types of abnormality but the system shows a stronger prediction power for type $S$. 

\begin{table}[t]
\footnotesize
\centering
\caption{predictive probability versus prior probability without windowing}
\label{table:pred}
\begin{tabular}{|c|l|l|l|l||l|l|l|}
\hline
\multicolumn{2}{|l|}{\multirow{2}{*}{}} & \multicolumn{3}{m{8em}|}{\# of predicted ground truth} & \multicolumn{3}{m{8em}|}{\% of predicted ground truth} \\ \cline{3-8} 
\multicolumn{2}{|l|}{}                  & V               & S               & F             & V               & S               & F             \\ \hline
\multirow{3}{2.5em}{\textit{yellow} alarm}    & V    & 467             & 122              & 14            & \textbf{77.45}  & 20.23           & 2.32          \\ \cline{2-8} 
                                 & S    & 36              & 15              & 0             & 70.59           & \textbf{28.41}  & 0             \\ \cline{2-8} 
                                 & F    & 40              & 60              & 5             & 38.10           & 57.14           & \textbf{4.76} \\ \hline
\multicolumn{2}{|c|}{total}             & 543             & 197             & 19            & \textbf{71.54}  & \textbf{25.96}  & \textbf{2.50} \\ \hline
\end{tabular}
\end{table}

In order to study the impact of the timing window (the time between the \textit{yellow} alarm and the subsequent \textit{red} alarm), we also applied a window of 10 consecutive samples following a \textit{yellow} alarm. Similarly, the prior and posterior probabilities are compared to evaluate the performance of the prediction capacity as shown in Table \ref{table:pred10}.

\begin{table}[t]
\footnotesize
\centering
\caption{predictive probability versus prior probability within 10 beats' window}
\label{table:pred10}
\begin{tabular}{|c|l|l|l|l||l|l|l|}
\hline
\multicolumn{2}{|l|}{\multirow{2}{*}{}} & \multicolumn{3}{m{8em}|}{\# of predicted ground truth} & \multicolumn{3}{m{8em}|}{\% of predicted ground truth} \\ \cline{3-8} 
\multicolumn{2}{|l|}{}                  & V               & S               & F             & V               & S               & F             \\ \hline
\multirow{3}{2.5em}{\textit{yellow} alarm}    & V    & 290             & 85              & 12            & \textbf{74.94}  & 21.96           & 3.10          \\ \cline{2-8} 
                                 & S    & 22              & 13              & 0             & 62.86           & \textbf{37.14}  & 0             \\ \cline{2-8} 
                                 & F    & 29              & 37              & 6             & 40.28           & 51.39           & \textbf{8.33} \\ \hline
\multicolumn{2}{|c|}{total}             & 341             & 135             & 18            & \textbf{69.03}  & \textbf{27.32}  & \textbf{3.64} \\ \hline
\end{tabular}
\end{table}

Compared with the result without windowing, the prediction performance within a 10-beat window shows that the proposed algorithm presents a better predictive power if a certain timing window is used. Especially for type S, the probability of observing a sample of type S within 10 beats after a \textit{yellow} alarm of type S is $37.14\%$, much higher than the prior probability of 27.32\%. This 10\% increase in the probability of observing an alarm of specific type provides a clear evidence that \textit{yellow} alarms are informative and should not be ignored. 

\section{Conclusion}\label{sec:conclusion}
We introduced a new framework for ECG signal processing. In contrast to the common practice, where the goal is to recognize severe heart abnormalities when they occur, we focused on in-depth analysis of seemingly normal signals. Our analysis verifies the previously report fact that normal samples may include mild morphology distortions that can be informative about upcoming severe abnormalities. This information is ignored in most existing classifiers. Increasing the sensitivity of current global classifiers is not an option, since it creates undesired annoying false alarms. 

Our two-stage analysis utilizes the concept of \textit{red} and \textit{yellow} alarms. A novel deviation analysis is developed, which builds and gradually refines a patient-specific reference for the patient's normal signal morphology. 
To facilitate the patient-specific deviation analysis, a novel spatial transformation is proposed to reshape the clustering geometry to achieve the desired \textit{symmetry} and \textit{separability} properties. 
\ifx \PSO \Yes
Furthermore, we proposed two \textit{PSO}-based numerical and \textit{orthogonalization-based} analytical solutions to optimize the transformation. 
\else
Furthermore, we proposed a deterministic optimization method based on \textit{orthogonalization} of abnormality cluster centroids to optimize the transformation. 
\fi
Applying this method to the benchmark ECG Dataset~\cite{mitdb} confirms that the classification accuracy is in the 95\% range which is comparable to (and in some cases is better than) the state of the art. 
More importantly, our system enables a unique feature of \textit{predictive analysis}. The results show that the probability of observing a \textit{red} alarm of specific type can be raised about 5\% to 10\% after calling a \textit{yellow} alarm of the same type. 
These results are consistent with the fact that most health problems evolve over time and often recognized when they become severe and present painful symptoms. 
This methodology can be used by health-care systems as well as wearable heart monitoring devices by elderly and high risk people to find out about potential upcoming heart problems and take precautions and preventive actions.
We finally note that this methodology is applicable to other biomedical signals, including electroencephalogram (EEG).

\bibliographystyle{model1-num-names}
\bibliography{Bibs}{}

\end{document}